\documentclass[11pt,a4paper,english,twoside]{article}

\usepackage{a4wide}
\usepackage{amssymb, amsmath, amsthm}
\usepackage{graphicx}
\usepackage{subcaption}
\usepackage[all]{xy}
\usepackage[pdftex,hyperref,svgnames]{xcolor}
\usepackage[pdftex,colorlinks=true,
pdfstartview=FitV,
pdfnewwindow=true,
linktoc = page,
linkcolor= blue,
citecolor= red,
urlcolor= blue,
hyperindex=true,
hyperfigures=false]{hyperref}
\hypersetup{linktocpage}
\usepackage{dsfont}
\usepackage{empheq}
\usepackage{cite}
\usepackage{float}
\usepackage{cancel}
\usepackage{relsize}
\usepackage{soul}
\usepackage{enumerate}
\usepackage{enumitem}
\usepackage{hhline}
\usepackage{multirow}
\usepackage{xspace}
\usepackage{bbm}
\usepackage{pdfpages}
\usepackage{setspace}
\usepackage{comment}

\newcommand{\nn}{\nonumber}

\def\beq{\begin{equation}}
\def\eeq{\end{equation}}
\def\bea{\begin{eqnarray}}
\def\eea{\end{eqnarray}}

\def\del {\partial}
\def\d {{\rm d}}

\begin{document}
\numberwithin{equation}{section}

\begin{titlepage}
\begin{center}

\phantom{DRAFT}

\vspace{1.8 cm}

{\LARGE \bf{Phantom matters}}\\

\vspace{2.0 cm} {\Large David Andriot}\\

\vspace{0.9 cm} {\small\slshape Laboratoire d'Annecy-le-Vieux de Physique Th\'eorique (LAPTh),\\
CNRS, Universit\'e Savoie Mont Blanc (USMB), UMR 5108,\\
9 Chemin de Bellevue, 74940 Annecy, France}\\

\vspace{0.9cm} {\upshape\ttfamily andriot@lapth.cnrs.fr}\\

\vspace{2.2cm}
{\bf Abstract}
\vspace{0.1cm}
\end{center}
\begin{quotation}

Cosmological observations of the recent universe suggest that dark energy equation of state parameter $w$ is growing with time, departing from a cosmological constant for which $w=-1$. Standard quintessence models allow for a varying $w\geq-1$, but observations report that a phantom regime, $w<-1$, is quickly reached in the past. Often discarded because of uncertainties or parametrisation, we rather propose here to embrace the reality of this phantom regime. We revisit an elegant mechanism that accounts for it, thanks to a coupling of quintessence field(s) to matter (and possibly radiation). We show that this allows for steep scalar potentials, and illustrate this with string-inspired models, where $V=V_0\, e^{-\lambda\, \varphi}$ and $\lambda \geq \sqrt{2}$. Those provide solutions in very good agreement with observations, including the phantom regime. We then discuss poles that can appear in $w$, making it diverge at recent times ($z\leq 4$), and that could be detected by observations. We finally comment on an Early Dark Energy-like feature, that systematically appears for free from the models considered, and could be of interest for the Hubble tension.

\end{quotation}

\end{titlepage}

\tableofcontents

\section{Introduction}

Dark energy is the constituent in our universe responsible for its current accelerated expansion. So far, observations have been compatible with the simplest form of dark energy, that is, a positive cosmological constant, $\Lambda>0$, thereby confirming the standard cosmological model $\Lambda$CDM. However, recent and on-going cosmological observations \cite{DES:2024jxu, DESI:2024mwx, DESI:2025zgx}, carried-out to unprecedented precision levels, suggest that other dark energy models could also be valid, namely models where constants are traded for evolving quantities. A simple realisation of such a dynamical dark energy is provided by quintessence models \cite{Ratra:1987rm, Peebles:1987ek, Wetterich:1994bg, Caldwell:1997ii}, where minimally coupled scalar field(s) $\{ \varphi^i \}$ evolve in a scalar potential $V$. The cosmological constant gets traded for the energy density $\rho_{\varphi} = {\rm kin} + V$, made up of the kinetic energy and the potential, both a priori non-constant. In addition, the dark energy equation of state parameter $w$, which is constant and $w=-1$ in the case of $\Lambda$, becomes as well an evolving quantity for quintessence
\beq
w_{\varphi} = \frac{{\rm kin} - V}{{\rm kin} + V} \ .
\eeq
In a standard situation where both ${\rm kin}$ and $V$ are positive and do not vanish simultaneously, it is straightforward to verify that
\beq
-1 \leq w_{\varphi}  \leq 1 \ .
\eeq
The dark energy equation of state parameter is obtained from observations. Recent experiments have successfully tested scenarios having a varying $w$, on various data sets between today (redshift $z=0$) and a recent past ($z=4$). More precisely, the model-independent CPL parametrisation was used, which consists in a linear expansion of $w$ in terms of the FLRW scale factor $a=1/(1+z)$,
\beq
w(a) = w_0 + w_a (1-a) \ ,
\eeq
where $a=1$ today. Values were obtained for the 2 constant parameters $w_0, w_a$. For instance, DESI + CMB + Union3 \cite{DESI:2025zgx} gives $w_0=-0.667 \pm 0.088$, $w_a=-1.09^{+0.31}_{-0.27}$, which excludes $\Lambda$CDM ($w_0=-1, w_a=0$) at an unprecedented precision.

\begin{figure}[t]
\begin{center}
\includegraphics[width=0.50\textwidth]{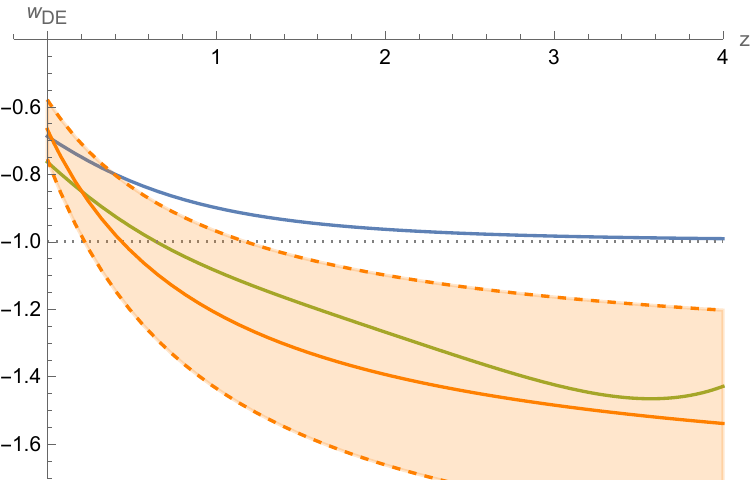}
\caption{Evolution of the dark energy equation of state parameter $w_{{\rm DE}}(z)$, along cosmological solutions in quintessence models with an exponential potential \eqref{Vexp}, for $\lambda =\sqrt{2}$. The first model (green) has a (string-inspired) coupling to matter, while the second one (blue) has no such coupling. Observational values (DESI + CMB + Union3) with CPL parametrisation are given in orange (plain central value and dashed error bars). While the minimally coupled model (blue) with such a steep stringy potential does not stand the comparison, the other model, having in addition a coupling to matter (green), is in very good agreement with observations.}\label{fig:ExpIntrowz}
\end{center}
\end{figure}

Given these observational results, a variety of quintessence models have been tested; we refrain from referring to this literature in detail here. It is for example the case of thawing models, for which $w$ goes from $-1$ during matter domination towards a higher value today; see e.g.~\cite{Andriot:2024sif} for a recent account. These models include exponential potentials
\beq
V = V_0 e^{-\lambda\, \varphi} \ , \label{Vexp}
\eeq
for $\lambda < \sqrt{6}$, where the single field $\varphi$ is canonically normalized, but also hilltop potentials ($V \sim 1- \kappa^2 \varphi^2$), axionic potentials ($V \sim 1+ \cos (\varphi/f)$), etc. The latter was directly compared to the data by the DESI collaboration \cite{DESI:2025fii}, and a string-inspired realisation in a transient regime was proposed in \cite{Anchordoqui:2025fgz}. Most string theory models however consider field space asymptotic, $\varphi \rightarrow \infty$, where corrections are controlled. In that case, $V$ is often exponential. In addition, $V$ is then subject to the Strong de Sitter Conjecture \cite{Rudelius:2021oaz, Bedroya:2019snp} from the swampland program, that suffers no counter-example up-to-date. This constraint requires $\lambda \geq \sqrt{2}$, i.e.~a steep potential. The case of an exponential potential has undergone many comparisons to the observational data, without \cite{Agrawal:2018own, Akrami:2018ylq, Raveri:2018ddi, Schoneberg:2023lun} or with \cite{Bhattacharya:2024hep, Alestas:2024gxe, Akrami:2025zlb} spatial curvature. The conclusion is that observations disagrees with a steep exponential potential, preferring $\lambda < 0.5 -1$. The single field asymptotic stringy potentials are therefore disfavored. This is illustrated in Figure \ref{fig:ExpIntrowz} by the blue curve ($\lambda=\sqrt{2}$) as compared to the observational orange curve.

There is however a (frightening) elephant in the room, with respect to the above studies of quintessence models: the {\sl phantom} regime. The latter refers to part of the universe history when $w < -1$. Having a phantom dark energy is in line with $w_0$ not too far from $-1$, and a slope $-w_a$ fairly steep (around $1$). It is also equivalent to $w_0 +w_a < -1 $ \cite{Andriot:2024sif}. Such values were already reported by the Planck collaboration \cite[Tab.6]{Planck:2018vyg}, with however larger error bars (see \cite{Ludwick:2017tox} for an early review on observations of a phantom regime and theoretical options; see also \cite{Escamilla:2023oce}). Similar values were also obtained recently by ACT \cite{ACT:2025tim}. Last but not least, the DES and DESI collaborations reported and emphasized such values. In addition, they indicated a very late (i.e.~close to today) phantom crossing $z_c$: backwards in time, $w$ is observed to diminish from $w_0$ towards $-1$, that is reached at $z_c \sim 0.5$. This means that most of the observed redshifts are in a phantom regime!

This situation is problematic for the above quintessence models, for which $w_{\varphi} \geq -1$. A first option is to ignore the phantom part, and only test the models on the most recent universe. This might be justified by the fact that errors bars on $w$ are more important (and possibly large) towards the past. However, the low $w_0$ and steep $w_a$ still make it difficult to match some of the quintessence models (see e.g.~\cite[Sec.4.4.3]{Andriot:2024sif}). Another option is to consider that this observed phantom regime is not real, but an artefact of the CPL parametrisation, that should not be used in the first place: see e.g.~\cite{Nesseris:2025lke, Shlivko:2025fgv, Akrami:2025zlb, Li:2025cxn, Wu:2025wyk} for a recent sample of tests of alternative parametrisations; see also \cite{DESI:2025fii} for related comments. In this paper, we propose to take an opposite view: not only do we embrace the reality of an observed phantom regime, but we also argue that it matters and should not be ignored, in the sense that this phantom dark energy teaches us important lessons on the model to be used.\\

Interestingly, alternative quintessence models have been proposed, that allow for a phantom regime. A first option is to consider negative kinetic terms \cite{Caldwell:1999ew}. This is along the lines of a violation of the Null Energy Condition (NEC), and various reasons (among which the difficulty to generate ${\rm kin}<0$ from string theory), make this option not satisfying in a healthy model. Further complicated theoretical ideas have been put forward (Horndeski and beyond, quantum spacetime discretization, massive gravity, etc.), a recent non-exhaustive sample being \cite{Tiwari:2024gzo, Oriti:2025lwx, Akarsu:2025gwi, Moghtaderi:2025cns, Ye:2025ulq, Smirnov:2025yru}. In this work, we rather focus on a simple and somewhat natural option, that is, allowing for a coupling $A_m(\varphi)$ of quintessence scalar field(s) to matter \cite{Amendola:1999qq, Billyard:2000bh}. It is indeed well-known that such a matter coupling generates a phantom dark energy \cite{Huey:2004qv, Das:2005yj}, through an elegant mechanism, reviewed in Section \ref{sec:mecha}.

In short, the mechanism amounts to identify matter as the energy density evolving as $a^{-3}$, as observations actually do. Any other source of evolution, such as the coupling function $A_m(\varphi)$, is then captured by the unknown dark energy density $\rho_{{\rm DE}}$. This effective definition of dark energy therefore includes the extra variation due to $A_m(\varphi)$, on top of the standard quintessence one from $\rho_{\varphi}$. An effective dark energy equation of state parameter, $w_{{\rm DE}}$, is similarly defined. It differs from $w_{\varphi}$ with a factor, that precisely allows $w_{{\rm DE}}$ to reach phantom values.

A time-varying coupling to matter is of course subject to many phenomenological constraints, which, to start with, would tend to restrict to a coupling to dark matter, and use e.g.~an axionic scalar field. Constraints include questions on dark matter mass variation, fifth and long-range forces, time variation of fundamental constants, etc. \cite{Uzan:2006mf, Uzan:2010pm, Uzan:2024ded}. Recent studies of such matter-coupled quintessence models include \cite{Ganesan:2024bsf, Khoury:2025txd, Chakraborty:2025syu, Smith:2025grk, Postolak:2025qmv}. It is also worth recalling, as reviewed in Section \ref{sec:gravrad}, that a matter coupling can equivalently be recast as a non-minimal coupling to gravity, by going from Einstein to Jordan frame. Constraints can thus be formulated there, and one can also obtain a phantom dark energy in such a framework \cite{Carvalho:2004ty, Martin:2005bp}. A recent study of such a model in view of the DESI phantom regime was carried-out in \cite{Wolf:2025jed}.\\

In this work, we consider a quintessence model with a coupling to matter (and possibly to radiation, as discussed in Section \ref{sec:gravrad}), and make use of the known mechanism to reach a phantom regime of dark energy. Both the potential and the coupling(s) considered are exponential. We are agnostic to the aforementioned phenomenological constraints. Our first aim is rather to emphasize the following point:
\bea
&& \text{{\sl In the described framework, having a phantom regime in the recent universe}}\nn\\
&& \text{{\sl requires a steeper potential $V$, when compared to having none.}}
\eea
In other words, having a phantom regime, realised through the mechanism described, actually calls for a steep potential. Consequently, the stringy asymptotic potentials, previously disfavored by observations, now actually stand a chance, thanks to the coupling to matter and phantom realisation. We illustrate this in Figure \ref{fig:ExpIntrowz}: considering models with a steep stringy potential ($\lambda=\sqrt{2}$), the one with matter-coupling agrees very well with observations, including in the phantom regime, contrary to the model without such coupling. In Section \ref{sec:string}, we present two examples of string-inspired models, with $\lambda \geq \sqrt{2}$ and string-inspired coupling functions. Those admit suitable cosmological solutions that appear in very good agreement with observations.

\begin{figure}[t]
\begin{center}
\begin{subfigure}[H]{0.48\textwidth}
\includegraphics[width=\textwidth]{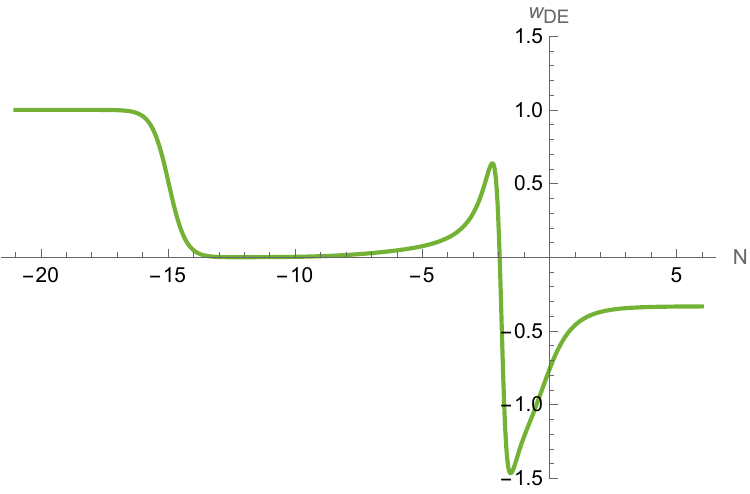}\caption{}\label{fig:ExpExp3wNbisintro}
\end{subfigure}\quad
\begin{subfigure}[H]{0.48\textwidth}
\includegraphics[width=\textwidth]{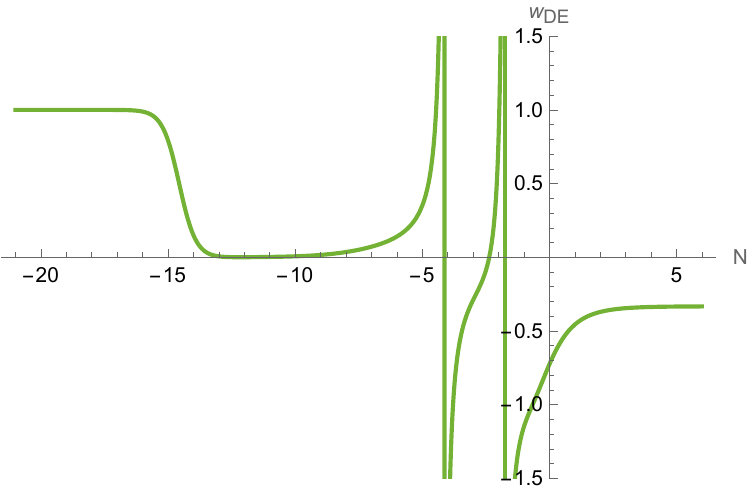}\caption{}\label{fig:ExpExp5wNintro}
\end{subfigure}
\caption{Evolution of $w_{{\rm DE}}(N)$ along suitable cosmological solutions, for models having an exponential potential \eqref{Vexp} with $\lambda=\sqrt{2}$, and a coupling to matter. The latter differs in one parameter by a factor of $2$ between the two models. This is enough to generate poles, as in Figure \ref{fig:ExpExp5wNintro}.}\label{fig:ExpExpwNintro}
\end{center}
\end{figure}

A second aim of this work is the study of two unphysical {\sl poles} that may appear in $w_{{\rm DE}}$, making it diverge, typically for $z\lesssim 4$. We describe those in Section \ref{sec:poles}, and show that their appearance is very sensitive to the choice of model parameters. We illustrate this in Figure \ref{fig:ExpExpwNintro}: two models differing by a factor of $2$ lead to poles (Figure \ref{fig:ExpExp5wNintro}) or not (Figure \ref{fig:ExpExp3wNbisintro}). We also comment on the current status of observational constraints in this regard, in particular whether $w_{{\rm DE}}(z)$ gets concave or convex. We argue that a thorough analysis in that respect should be within reach, and would be very informative on the details of an adequate model (potential and coupling function).

We finally provide in Section \ref{sec:out} a brief summary and an outlook. We make the observation that a bump in $\Omega_{{\rm DE}}$ appears around $z \approx 10^3$, that is very reminiscent of Early Dark Energy models. We understand why this feature is systematic in the solutions considered, and therefore comes out for free. Evaluating this effect with respect to the Hubble tension is beyond the scope of this work, but this feature is appealing.

While models considered in this work involve an exponential potential and exponential couplings, we introduce in Appendix \ref{ap:hilltop} a different one, with a hilltop potential and a linear coupling. We present a cosmological solution and discuss its observational validity.

\section{Framework}\label{sec:form}

In this section, we present the cosmological models to be used, and review how those can give rise to a phantom regime for an effective dark energy component; we follow partially \cite{Das:2005yj, Khoury:2025txd}. We discuss the coupling of a quintessence field to matter, but also to gravitation and to radiation.

\subsection{Mechanism to get a phantom regime from matter coupling}\label{sec:mecha}

We consider the following 4-dimensional (4d) cosmological model
\beq
\int \d^4 x \sqrt{|g_4|} \left( \frac{M_p^2}{2} {\cal R}_4 - \frac{1}{2} g_{ij} \del_{\mu} \varphi^i \del^{\mu} \varphi^j - V(\varphi^k) - L_m - L_r \right) \ , \label{action}
\eeq
following notations of \cite{Andriot:2024sif}. This gravitational model is expressed in Einstein frame, with scalar fields $\varphi^i$ minimally coupled to gravity, and with a field space metric $g_{ij}$, a scalar potential $V$, and matter and radiation Lagrangian densities $L_m$ and $L_r$. In this work, we allow the latter two to depend on some scalars $\varphi^k$, in other words the scalar fields can have a coupling to matter and/or radiation.

One can derive as usual the Einstein equations. Using an FLRW ansatz for the metric, with signature $(-,+,+,+)$, scale factor $a(t)$, and taking here for simplicity a spatially flat universe ($k=0$), one obtains the standard two Friedmann equations $F_1=F_2=0$, where
\beq
F_1 = 3 H^2 - \frac{1}{M_p^2} \sum_n \rho_n \ ,\quad F_2 = \dot{H} + \frac{1}{2 M_p^2} \sum_n (1+w_n) \rho_n \ .
\eeq
We denoted the Hubble parameter $H=\dot{a}/a$, with the dot standing for the time derivative. The label $n$ corresponds to the various constituents, $n= \varphi,m,r$, appearing through their energy densities $\rho_n$ and equation of state parameter $w_n$. The $\rho_n$ are obtained from the action \eqref{action} via the energy momentum tensor. We refer to \cite{Andriot:2024sif} for further details.

In the following, we will restrict to fields being only time dependent, $\varphi^i(t)$, and set the 4d reduced Planck mass $M_p=1$. We will also restrict to $V\geq 0$, and to a definite positive field space metric $g_{ij}$, giving a positive kinetic energy $1/2 (\dot{\varphi})^2 \equiv 1/2 g_{ij} \dot{\varphi}^i \dot{\varphi}^j \geq 0$. We also assume that the two contributions never vanish together, resulting in $\rho_{\varphi}>0$, as well as a bounded equation of state parameter
\beq
w_{\varphi} = \frac{\frac12 (\dot{\varphi})^2  -V}{\frac12 (\dot{\varphi})^2  + V} \ ,\quad -1 \leq w_{\varphi} \leq 1 \ .
\eeq

We consider a coupling of scalar fields to matter via a multiplicative function $A_m(\varphi^k) >0$: $L_m (\varphi^k, g_{\mu\nu}, \psi) = A_m(\varphi^k) \, \bar{L}_m (g_{\mu\nu}, \psi)$. This results in the following energy density definition
\beq
\rho_m = A_m(\varphi^k) \, \bar{\rho}_m \ ,\ {\rm with}\ \bar{\rho}_m = \rho_{m0} \, a^{-3} \ .\label{rhom}
\eeq
Here and in the following, we denote by the subscript ${}_0$ values of quantities today; we take in particular $a_0 =1$. Without loss of generality, we also choose the value today $A_m(\varphi^k_0) = 1$. This coupling induces a change in the field equations $E^i =0$:\footnote{This derivation implicitly requires the on-shell relation $\bar{L}_m = \bar{\rho}_m$. This is consistent with the energy-momentum tensor, provided that $\bar{L}_m$ is inversely proportional to the spatial volume, $\sqrt{g_{11} g_{22} g_{33}}$, as expected.} those are now given by
\beq
E^i = \ddot{\varphi}^i + \Gamma^i_{jk} \dot{\varphi}^j \dot{\varphi}^k + 3 H \dot{\varphi}^i + g^{ij} \left( \del_{\varphi^j} V + \bar{\rho}_m\, \del_{\varphi^j} A_m  \right) \ , \label{Ei}
\eeq
where $A_m$ or $\rho_m$ provides a new, potential-like, contribution, giving overall an effectively modified potential. $\Gamma^i_{jk}$ stands for the Christoffel symbol for $g_{ij}$.\\

{\sl Observationally}, matter is identified as a density that varies as $a^{-3}$; similarly, radiation is identified as a variation in $a^{-4}$, and curvature as $a^{-2}$, even though those two are either not considered or negligible in the recent universe. {\sl Any other evolving component will then be identified, observationally, as dark energy.} This gives rise to the effective definition of dark energy, that we label ${}_{{\rm DE}}$ in the following. In particular here, the variation of $\rho_m$ due to $A_m$ would be captured by the effective dark energy density, contributing to it on top of the purely quintessence scalar field part. We thus define the effective dark energy density, $\rho_{{\rm DE}}$, as
\beq
\rho_m + \rho_{\varphi} = \bar{\rho}_m + \rho_{{\rm DE}} \ \Rightarrow \ \rho_{{\rm DE}} \equiv \rho_{\varphi} + (A_m-1) \bar{\rho}_m \ .
\eeq
The equation $F_1=0$ and its dynamics are then preserved, as we can now run $n$ on the new constituents ($\rho_r, \bar{\rho}_m,  \rho_{{\rm DE}}$). Similarly, the effective equation of state parameter is defined in such a way that $F_2$ is preserved,\footnote{\label{foot:wm}The energy density and pressure are defined as components of the energy-momentum tensor $T_{\mu\nu}$, which is obtained from the Lagrangian variation with respect to the metric. Regarding matter, having or not in the Lagrangian a multiplicative coupling function $A_m$, depending on the independent fields $\varphi^i$, does not affect the derivation of $T_{\mu\nu}$: $A_m$ also appears multiplicatively in the resulting energy density and pressure, as in \eqref{rhom}. As a consequence, the matter equation of state parameter is unchanged: $w_m= \bar{w}_m=0$. The same reasoning can be made for radiation, in case of a coupling there.} namely
\beq
w_{\varphi}\, \rho_{\varphi} = w_{{\rm DE}}\, \rho_{{\rm DE}} \ \Rightarrow \ w_{{\rm DE}} \equiv w_{\varphi} \frac{\rho_{\varphi} }{\rho_{{\rm DE}}} = \frac{w_{\varphi}}{1 + (A_m-1) \frac{\bar{\rho}_m}{\rho_{\varphi}} } \ . \label{wDE}
\eeq

We now verify that these definitions give the expected continuity equation for the effective dark energy. First, one has
\beq
\dot{\rho}_{\varphi} = \dot{\varphi}^i g_{ij} \left( E^j - 3 H \dot{\varphi}^i - g^{jk} \bar{\rho}_m\, \del_{\varphi^k} A_m  \right) \ .
\eeq
With $E^j=0$, this gets rewritten as
\beq
\dot{\rho}_{\varphi} =  - 3 H (1 +w_{\varphi}) \rho_{\varphi} - \bar{\rho}_m\, \dot{A}_m \ .
\eeq
From the above definitions, it is then straightforward to verify the expected continuity equation
\beq
\dot{\rho}_{{\rm DE}} = - 3 H (1+ w_{{\rm DE}} )\, \rho_{{\rm DE}} \ ,
\eeq
which is used for observations. Let us note that in ``models of interacting dark energy'', the matter and dark energy continuity equations are phenomenologically modified (see e.g.~\cite[(2),(3)]{Silva:2025hxw}). Here, while equations for $\rho_m, \rho_{\varphi}$ get indeed modified, we argued that those for the observed densities $\bar{\rho}_m, \rho_{{\rm DE}}$ are unchanged: this is an important distinction when setting observational constraints.

All conventions are such that the values for the various constituents match today. For instance, introducing the energy density parameters $\Omega_n= \frac{\rho_n}{3 H^2}$, one has
\beq
\bar{\Omega}_{m0} = \Omega_{m0} \ , \ \Omega_{{\rm DE}0} =\Omega_{\varphi0} \ ,\ w_{{\rm DE}0} =w_{\varphi0} \ .
\eeq
However, while we explained that $w_{\varphi} \geq -1$, the effective dark energy can reach a phantom regime, namely
\beq
w_{{\rm DE}} < -1 \ .
\eeq
From the definition \eqref{wDE}, we see that a necessary condition to achieve this is that the denominator gets smaller than $1$, requiring
\beq
A_m \leq 1 \ .
\eeq
To get a phantom regime in the recent (past) universe, we deduce that $A_m$ has to be growing with time, at least for some time range. This brings us to one of the main points of this paper.

Let us consider a quintessence model where in the recent universe, the fields are rolling down a potential, namely $\dot{\varphi}^i> 0$ with $\del_{\varphi^i} V < 0$. This is a natural and expected evolution in a realistic quintessence scenario. Indeed, the thawing models (see \cite{Andriot:2024sif} for a recent account) have mostly frozen fields during radiation and matter domination due to a high Hubble friction, leading to fields rolling down the potential in the recent universe when dark energy starts becoming relevant. Such a thawing evolution provides an increasing, not decreasing, equation of state parameter $w_{\varphi}$, in agreement with recent observation which have $w_a<0$. We thus place ourselves in such a situation for the recent universe. In that case, having $A_m$ growing with time gets translated into
\beq
\del_{\varphi^i} A_m > 0 \ .
\eeq
In the field equations \eqref{Ei}, this contribution thus balances $\del_{\varphi^i} V $. We conclude
\bea
&& \text{{\sl In the described framework, having a phantom regime in the recent universe}}\nn\\
&&\text{{\sl requires a steeper potential $V$, when compared to having none.}}
\eea
As explained in the Introduction, string theory constructions in their asymptotic regimes would allow for steep exponential potentials, which are disfavored by observations when one restricts those to a non-phantom regime. Considering the phantom regime thus gives those potentials a chance to be observationally valid; we will verify this in the next section. Prior to this, let us extend the above to other possible couplings.

\subsection{Gravitational or radiation coupling}\label{sec:gravrad}

It is well-known that coupling to matter $ A_m (\varphi^k)$ as considered above, in 4d Einstein frame can be traded for a non-minimal coupling to gravity in the so-called Jordan frame (see e.g.~\cite{Pettorino:2008ez}): schematically, we denote this relation as follows at the level of the Lagrangian
\bea
&& \sqrt{|g_4|} \left( \frac{M_p^2}{2} {\cal R}_4  - A_m\, \bar{L}_m (g_{\mu\nu}, \psi) \right) \ \leftrightarrow \ \sqrt{|g_{J4}|} \left( \frac{M_p^2}{2} A_m^{-2}\, {\cal R}_{J4}  - \bar{L}_m (g_{J\,\mu\nu}, \psi) \right) \\
&& {\rm with}\ g_{J\,\mu\nu} = A_m^2\, g_{\mu\nu} \nn \ .
\eea
This change of frame also requires scalar field redefinitions to have canonically normalized ones, and a rescaling of the scalar potential. This equivalence can also be derived at the level of equations of motion, as e.g.~in \cite{Amendola:1999qq}, where one can use the energy density of matter and its scaling with $a$; the latter being maybe clearer than the generic metric dependence of $L_m$. Eventually, the equations of motion of both Lagrangians are shown to be mapped by this conformal transformation of the metric.

It is also well-known that the frame transformation, from Einstein to Jordan, does not change the radiation contribution, which is said to be conformally invariant. This can be seen directly at the level of the Lagrangian (but also through the equations of motion and $\rho_r$): indeed, it is straightforward to verify the invariance of $\sqrt{|g_4|} F_{\mu\nu} F^{\mu\nu}$. This might be a reason why a coupling to radiation is less often considered than one to matter.\\

Motivated by string theory or more generally dimensional reductions from theories with extra dimensions, we now consider a simple mechanism that would generate a coupling to radiation. Similarly to the above, it amounts to a change of frame, from what one may call a 4d string frame, towards the 4d Einstein frame. In the former, the Lagrangian is given by
\beq
\sqrt{|g_{S4}|}\, A\, \left( \frac{M_p^2}{2} {\cal R}_{S4}  - \bar{L}_m (g_{S\,\mu\nu}) - \bar{L}_r (g_{S\,\mu\nu}) \right) \ ,
\eeq
with the function of interest being $A$. The motivation for having such an overall factor comes from higher dimensional theories, from having the volume factor for the extra dimensions. With such an origin, the coupling constants in the matter and radiation Lagrangians could also depend on the extra dimensional metric; we ignore such a possibility here for simplicity. To perform the conformal transformation of the 4d metric and change of frame, we use the same dependence of $\bar{L}_{m,r}$ as described above, while one may equivalently derive this equivalence at the level of equations of motion. We eventually obtain
\beq
\sqrt{|g_{4}|} \left( \frac{M_p^2}{2} {\cal R}_{4}  - A^{\frac{1}{2}} \, \bar{L}_m (g_{\mu\nu}) - A\, \bar{L}_r (g_{\mu\nu}) \right) \ ,\ {\rm with}\ g_{S\,\mu\nu} = A^{-1}\, g_{\mu\nu} \ ,
\eeq
giving a coupling to radiation in addition to one to matter.\\

This motivates us to generalize the above definitions of an effective dark energy. For completeness, we now treat the two couplings independently, and introduce a new function $A_r$ on top of $A_m$ in the Lagrangian. We then define
\beq
\rho_m = A_m(\varphi^k) \, \bar{\rho}_m \ ,\ {\rm with}\ \bar{\rho}_m = \rho_{m0} \, a^{-3} \ ,\quad \rho_r = A_r(\varphi^k) \, \bar{\rho}_r \ ,\ {\rm with}\ \bar{\rho}_r = \rho_{r0} \, a^{-4} \ ,
\eeq
with $A_r(\varphi^k_0) = A_m(\varphi^k_0) = 1$. Dark energy density is effectively given by
\beq
\rho_r + \rho_m + \rho_{\varphi} = \bar{\rho}_r + \bar{\rho}_m + \rho_{{\rm DE}} \ \Rightarrow \ \rho_{{\rm DE}} \equiv \rho_{\varphi} + (A_m-1) \bar{\rho}_m  + (A_r-1) \bar{\rho}_r \ ,
\eeq
preserving $F_1=0$, and
\bea
\frac{1}{3} \, \rho_r + w_{\varphi}\, \rho_{\varphi} = \frac{1}{3} \, \bar{\rho}_r + w_{{\rm DE}}\, \rho_{{\rm DE}} \ \Rightarrow \ w_{{\rm DE}} &\equiv&  \frac{ w_{\varphi}\, \rho_{\varphi} + \frac{1}{3} \, (A_r-1)\bar{\rho}_r }{\rho_{{\rm DE}}} \label{wDEmr}\\
&=& \frac{w_{\varphi} + \frac{1}{3} \, (A_r-1) \frac{\bar{\rho}_r}{\rho_{\varphi}} }{1 + (A_m-1) \frac{\bar{\rho}_m}{\rho_{\varphi}} + (A_r-1) \frac{\bar{\rho}_r}{\rho_{\varphi}} } \ , \nn
\eea
preserving $F_2=0$ (see Footnote \ref{foot:wm}). The field equation $E^i=0$ is modified towards
\beq
E^i = \ddot{\varphi}^i + \Gamma^i_{jk} \dot{\varphi}^j \dot{\varphi}^k + 3 H \dot{\varphi}^i + g^{ij} \left( \del_{\varphi^j} V + \bar{\rho}_m\, \del_{\varphi^j} A_m  + \bar{\rho}_r\, \del_{\varphi^j} A_r  \right) \ ,
\eeq
giving a new effective potential.\footnote{As for the matter, the derivation of the radiation contribution requires here the on-shell matching $\bar{L}_r = \bar{\rho}_r$. This is consistent with the energy-momentum tensor, provided $\bar{L}_r$ is independent of $g^{00}$. With Maxwell Lagrangian, the latter restriction amounts to have no electric contribution.} From $E^j=0$, we get
\beq
\dot{\rho}_{\varphi} =  - 3 H (1 +w_{\varphi}) \rho_{\varphi} - \bar{\rho}_m\, \dot{A}_m - \bar{\rho}_r\, \dot{A}_r\ ,
\eeq
from which we can reproduce consistently the continuity equation for the effective dark energy
\beq
\dot{\rho}_{{\rm DE}} = - 3 H (1+ w_{{\rm DE}} )\, \rho_{{\rm DE}} \ .
\eeq
As above, having $A_{m,r}\leq 1$ and growing in the recent universe can help to get a phantom dark energy, which in turn allows for a {\sl steeper} $V$. We will provide examples of such solutions.

\section{Realistic cosmological solutions from string-inspired models}\label{sec:string}

In this section, we present cosmological solutions obtained from string-inspired models in the framework presented above. Our main point is to show that steep potentials, which could be argued to have a string theory origin, provide solutions in good agreement with observational constraints. This is achieved thanks to the mechanism described previously, where coupling to matter (and possibly radiation) allows to have a phantom regime for dark energy in the recent universe, as well as to accommodate a steep potential.\\

Let us first review the mathematical problem that will be numerically tackled to get the solutions. With the coupling to matter and radiation $A_{m,r}$ introduced above, one verifies that the following relation between equations is still obeyed
\beq
\dot{F}_1 = -\dot{\varphi}^i g_{ij} E^j + 6 H F_2 \ . \label{releq}
\eeq
This allows us to solve only $E^i=F_2=0$. $F_1$ is then a constant, which vanishes as it should once we pick the fiducial values today
\beq
\Omega_{\varphi0}=0.6850 \ ,\ \Omega_{m0}=0.3149 \ ,\ \Omega_{r0}=0.0001 \ .
\eeq
Extending \cite[(2.20)]{Andriot:2024sif}, the equations to solve get rewritten as follows, where we restrict in our examples to a single canonically normalized field
\bea
&&\hspace{-0.4in} 2 \tilde{H}\del_N \tilde{H} + 4 A_r(\varphi)\, \Omega_{r0}\, e^{-4N} + 3 A_m(\varphi)\, \Omega_{m0}\, e^{-3N} + \tilde{H}^2 (\del_N \varphi)^2 = 0 \ ,\\
&&\hspace{-0.4in} \tilde{H}^2 \del_N^2 \varphi + \tilde{H} \del_N \tilde{H} \, \del_N \varphi + 3 \tilde{H}^2 \del_N \varphi + \del_\varphi \tilde{V} + 3 \Omega_{m0}\, e^{-3N}\, \del_{\varphi} A_m  + 3 \Omega_{r0}\, e^{-4N}\, \del_{\varphi} A_r = 0 \ .
\eea
We included both couplings to matter and radiation $A_{m,r}$ for completeness. We introduced the notations $\tilde{V} = \frac{V}{H_0^2}, \tilde{H}= \frac{H}{H_0}$. To solve the problem and find a solution, one is left to specify $\tilde{V}, A_{m,r}$, and $w_{\varphi0}$. For the initial conditions, one takes \cite{Andriot:2024sif} $\tilde{H}_0=1$ and
\beq
\del_N \varphi|_0 = \sqrt{3 \Omega_{\varphi0}(1+w_{\varphi0})} \ ,\quad \tilde{V}|_0=\frac{3}{2} \Omega_{\varphi0}(1-w_{\varphi0}) \ , \label{IC}
\eeq
where $\tilde{V}|_0$ fixes either a potential constant or $\varphi_0$. As explained in \cite{Andriot:2024sif}, the value of $w_{\varphi0}$ gets fixed very precisely (i.e.~as a fine tuning) by picking a solution that would allow in the past to have not only matter but also radiation domination. We make the radiation domination start, in terms of e-folds, around $N=-20$, which corresponds to BBN.

In the following, we will consider an exponential potential, which is standard from string theory in the field space asymptotic, where corrections are under control
\beq
V=V_0\, e^{-\lambda \, \varphi} \ .
\eeq
For a string theory origin, we take $\lambda \geq \sqrt{2}$. There is a degeneracy between $V_0$ and a redefinition of the field by a shift: we can then fix one or the other. We decide to pick $\varphi_0=0$, and fix the constant $V_0$, or rather $\tilde{V}_0$, by the value of the potential today as in \eqref{IC}.

We are left to specify the couplings $A_{m,r}$. Without a detailed string theory construction providing particle physics (see e.g.~\cite{Marchesano:2024gul}), it is meaningless to argue too precisely on the form of these functions. We will still suggest natural, string-inspired choices in the following. As we will see in Section \ref{sec:poles}, more accurate observational constraints would also be necessary to be decisive on the details of these functions. The following is thus an illustration of the stringy possibilities for these couplings.\\

The first stringy example is inspired by the distance conjecture \cite{Ooguri:2006in}. For an evolving scalar field $\varphi$, the conjecture claims the existence of a tower of massive states, that could be interpreted e.g.~as dark matter. The mass of these states evolves as $m \sim e^{-\alpha \varphi}$, where in Calabi-Yau compactifications, $\alpha \geq \frac{1}{2} \sqrt{\frac{2}{3}}$ \cite{Grimm:2018ohb, Andriot:2020lea}. A scalar energy density would go as $m^2$, so we consider a coupling function $A_m$ that goes as $e^{-\sqrt{\frac{2}{3}} \varphi}$ (see e.g.~\cite{Agrawal:2019dlm, Gonzalo:2022jac, Casas:2024oak} for related works and ideas). Unfortunately, this is a decreasing function, while we argued in favor of an increasing one. We then extend our coupling function ansatz as follows (with proper normalization for today's value)
\beq
A_m(\varphi) = 1+A_{m\,\varphi} (1 - e^{-\alpha_m \varphi}) \ .
\eeq
One may also argue that having a constant term in $A_m$ allows to have only part of the matter (or even dark matter) being coupled.\footnote{For example, one may split non-relativistic matter in two pieces, $\bar{\rho}_m = \bar{\rho}_{{\rm B}} + \bar{\rho}_{{\rm DM}}$, both evolving as $a^{-3}$, and only couple the quintessence field to the second part: $\rho_m = \bar{\rho}_{{\rm B}} + \alpha(\varphi) \bar{\rho}_{{\rm DM}}$, with $\alpha(\varphi_0)=1$. In that case, one obtains $A_m(\varphi)= 1 +\tfrac{\bar{\rho}_{{\rm DM}0}}{\bar{\rho}_{m0}} (\alpha(\varphi) - 1)$, exhibiting a constant term.} As mentioned above, we refrain at this stage from discussing in more detail the function $A_m$. We consider in this example no radiation coupling, $A_r=0$.

\begin{figure}[t]
\begin{center}
\begin{subfigure}[H]{0.48\textwidth}
\includegraphics[width=\textwidth]{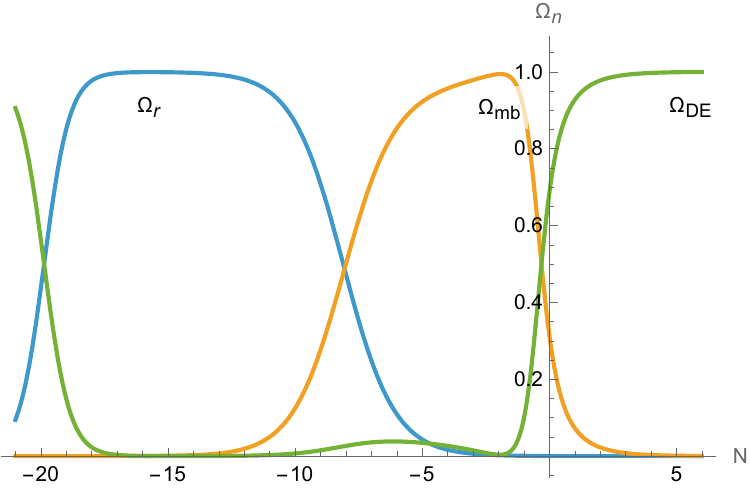}\caption{}\label{fig:ExpExp3Os}
\end{subfigure}\quad
\begin{subfigure}[H]{0.48\textwidth}
\includegraphics[width=\textwidth]{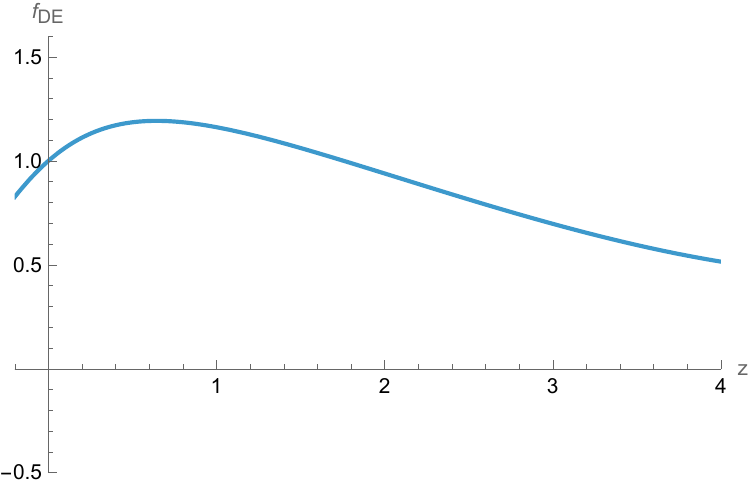}\caption{}\label{fig:ExpExp3fDEz}
\end{subfigure}\\
\begin{subfigure}[H]{0.48\textwidth}
\includegraphics[width=\textwidth]{figExpExp3wN.pdf}\caption{}\label{fig:ExpExp3wN}
\end{subfigure}\quad
\begin{subfigure}[H]{0.48\textwidth}
\includegraphics[width=\textwidth]{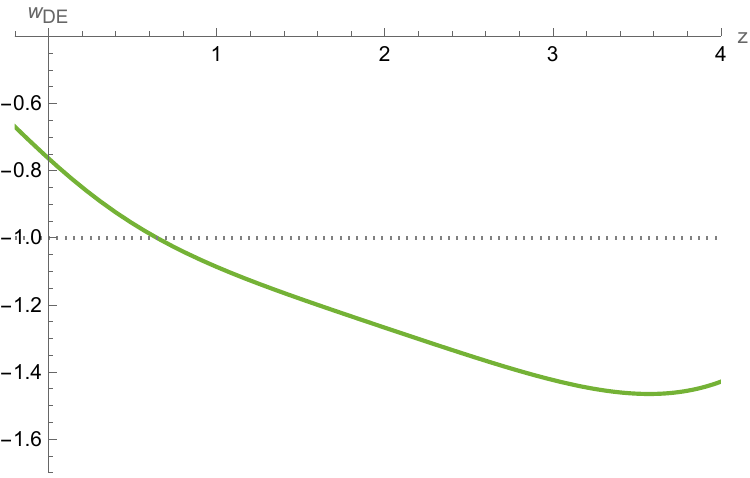}\caption{}\label{fig:ExpExp3wz}
\end{subfigure}
\caption{Cosmological solution in the string-inspired model with exponential potential and exponential coupling, as given in \eqref{solstring1}. The evolution of the $\Omega_n$ in terms of e-folds $N=\log a$ is provided in Figure \ref{fig:ExpExp3Os}, where $\Omega_{mb}$ stands for $\bar{\Omega}_m$. Figure \ref{fig:ExpExp3fDEz} features the evolution of the effective dark energy density, $f_{{\rm DE}}=\frac{\rho_{{\rm DE}}}{\rho_{{\rm DE}0}}$, in terms of redshift $z$. Figure \ref{fig:ExpExp3wN} and \ref{fig:ExpExp3wz} display the evolution of the effective dark energy equation of state parameter $w_{{\rm DE}}$.}\label{fig:ExpExp3}
\end{center}
\end{figure}

In short, we take the following string-inspired parameter values
\beq
\lambda=\sqrt{2}\ ,\ \alpha_m=\sqrt{\frac{2}{3}}\ ,\  A_{m\,\varphi}= \frac{1}{8}\ ,\quad {\rm and}\ w_{\varphi0}=-0.76365999931 \ ,\label{solstring1}
\eeq
giving the solution displayed in Figure \ref{fig:ExpExp3}.

We see in Figure \ref{fig:ExpExp3} a fairly standard evolution of the  $\Omega_n$, starting with a hypothetical kination phase as in thawing models, and followed by the usual radiation, matter and dark energy domination phases. We then see the realisation of the mechanism described previously: the dark energy admits a phantom regime in the recent universe, achieving phantom crossing ($w_{{\rm DE}}=-1$) at  $z_c =0.64$. It is easy to see, from the continuity equation ($\dot{\rho}_{{\rm DE}} \propto -(1+w_{{\rm DE}})$), that the phantom crossing corresponds to a maximum of $\rho_{{\rm DE}}$, as verified in Figure \ref{fig:ExpExp3fDEz}.

\begin{figure}[H]
\begin{center}
\begin{subfigure}[H]{0.48\textwidth}
\includegraphics[width=\textwidth]{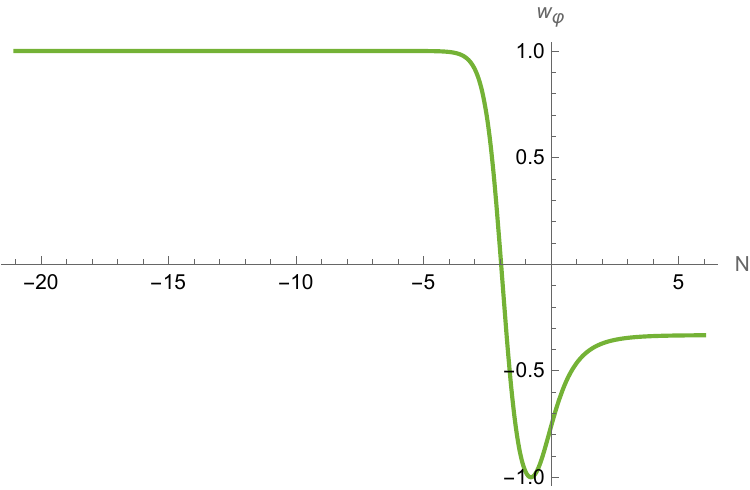}\caption{}\label{fig:ExpExp3wphiN}
\end{subfigure}\quad
\begin{subfigure}[H]{0.48\textwidth}
\includegraphics[width=\textwidth]{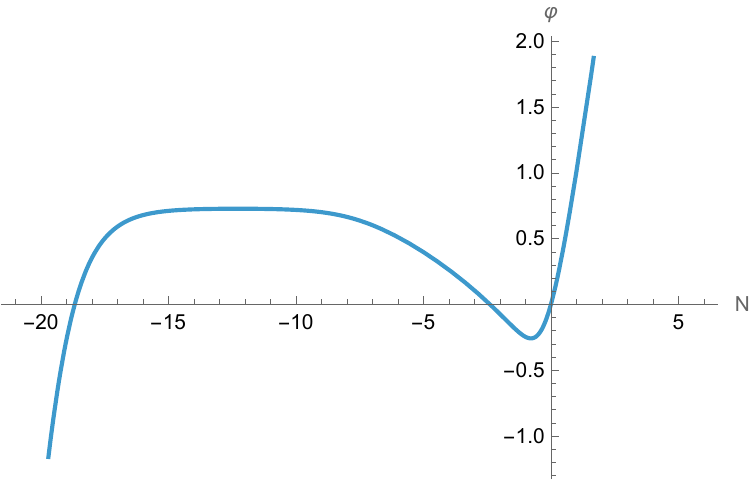}\caption{}\label{fig:ExpExp3phiN}
\end{subfigure}
\caption{Evolution of $w_{\varphi}(N)$ and $\varphi(N)$ in the cosmological solution corresponding to \eqref{solstring1} and Figure \ref{fig:ExpExp3}. There is one point, around $N\approx -12.28$, where the kinetic energy vanishes, but it grows very quickly on both sides of this point, due to a large $\tilde{H}$, in such a way that we do not notice this point in $w_{\varphi}(N)$.}\label{fig:ExpExp3wp}
\end{center}
\end{figure}

For completeness, we provide in Figure \ref{fig:ExpExp3wp} further elements of this interesting solution. The non-trivial evolution of $\varphi(N)$ underlines the role of $A_m$ in the effective potential, beyond $V$.

\begin{figure}[H]
\begin{center}
\begin{subfigure}[H]{0.48\textwidth}
\includegraphics[width=\textwidth]{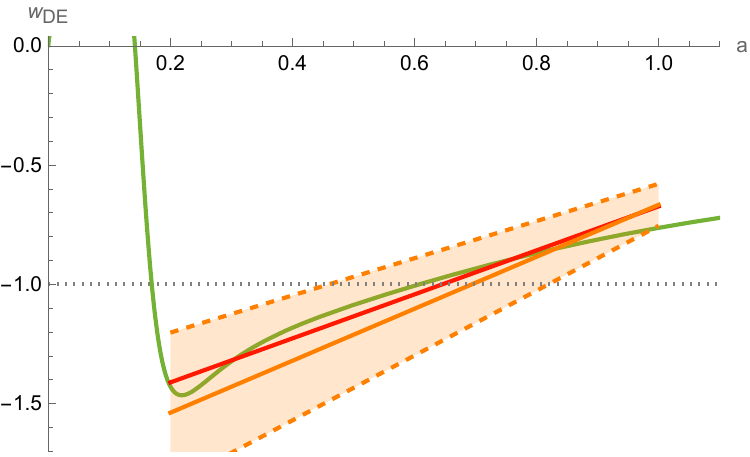}\caption{}\label{fig:ExpExp3wafit}
\end{subfigure}\quad
\begin{subfigure}[H]{0.48\textwidth}
\includegraphics[width=\textwidth]{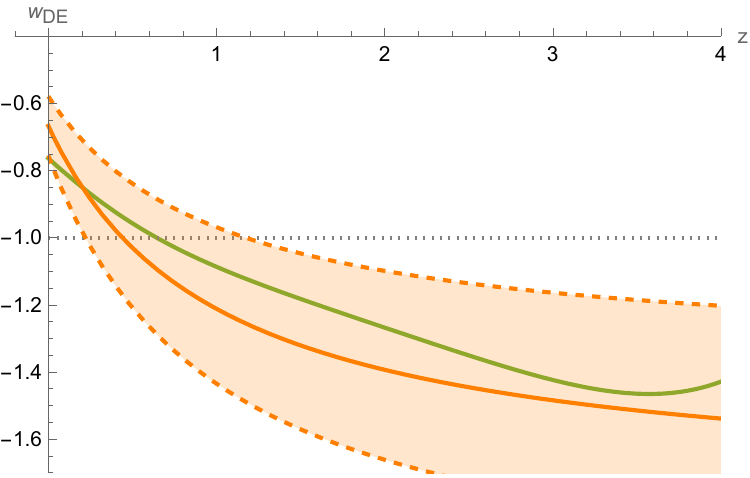}\caption{}\label{fig:ExpExp3wzfit}
\end{subfigure}
\caption{Comparison of the solution of Figure \ref{fig:ExpExp3} to the observational constraints from DESI + CMB + Union3 (DR2) \cite{DESI:2025zgx}. Figure \ref{fig:ExpExp3wafit} displays $w_{{\rm DE}}(a)$ in green with its linear fit in red, while Figure \ref{fig:ExpExp3wzfit} displays $w_{{\rm DE}}(z)$ in green. The observational values (central and error bars), based on CPL parametrisation, are displayed in orange, for $0\leq z \leq 4$.}\label{fig:ExpExp3fit}
\end{center}
\end{figure}

We now show that this solution is in very good agreement with observations. A linear fit of $w_{{\rm DE}}(a)$ to match the CPL parametrisation $w(a) =w_0 + w_a (1-a)$ gives
\beq
w_0 = -0.67\ ,\ w_a = -0.92 \ ,
\eeq
which agrees with observational values obtained from DESI + CMB + Union3 (DR2) \cite{DESI:2025zgx}. We display this fit, and the excellent agreement to observational constraints, including in the phantom regime, in Figure \ref{fig:ExpExp3fit}.

The second string-inspired example builds on the 6d volume dependence discussed in Section \ref{sec:gravrad}. There it was argued that an overall function $\sqrt{|g_6|} \sim A(\varphi)$ would eventually lead in 4d Einstein frame to a matter coupling $A_m \sim A^{\frac{1}{2}}$ as well as a radiation one $A_r \sim A$. We now consider this volume to be given by a canonical field $\varphi$, namely (from 10d Einstein frame) $\sqrt{|g_6|} \sim e^{\sqrt{\frac{3}{2}} \varphi}$. We then consider the case of a curvature generated potential (see e.g.~\cite{Andriot:2023wvg} or \cite[Foot.8]{Andriot:2025cyi}), as obtained from a compactification on a negatively curved compact Einstein manifold. Such a stringy setting gives rise to the following dependences
\beq
V \sim e^{- \sqrt{\frac{8}{3}} \varphi} \ ,\ A_r \sim e^{\sqrt{\frac{3}{2}} \varphi} \ ,\ A_m \sim e^{\sqrt{\frac{3}{8}} \varphi} \ .
\eeq
The interest of such a setting is that the coupling functions are growing, while the potential is decreasing, as desired for the above mechanism to work.

\begin{figure}[H]
\begin{center}
\begin{subfigure}[H]{0.48\textwidth}
\includegraphics[width=\textwidth]{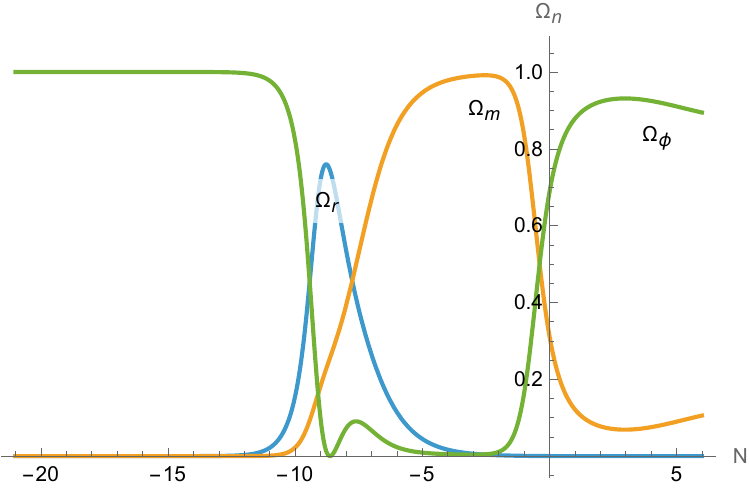}\caption{}\label{fig:ExpExpradOps}
\end{subfigure}\quad
\begin{subfigure}[H]{0.48\textwidth}
\includegraphics[width=\textwidth]{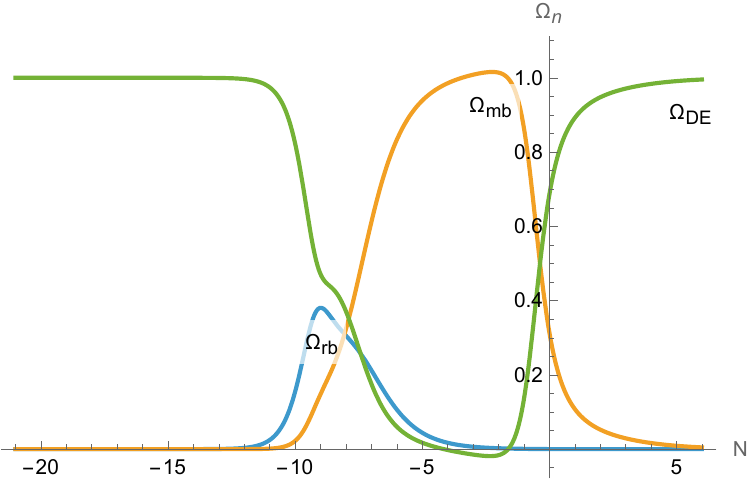}\caption{}\label{fig:ExpExpradOs}
\end{subfigure}\\
\begin{subfigure}[H]{0.48\textwidth}
\includegraphics[width=\textwidth]{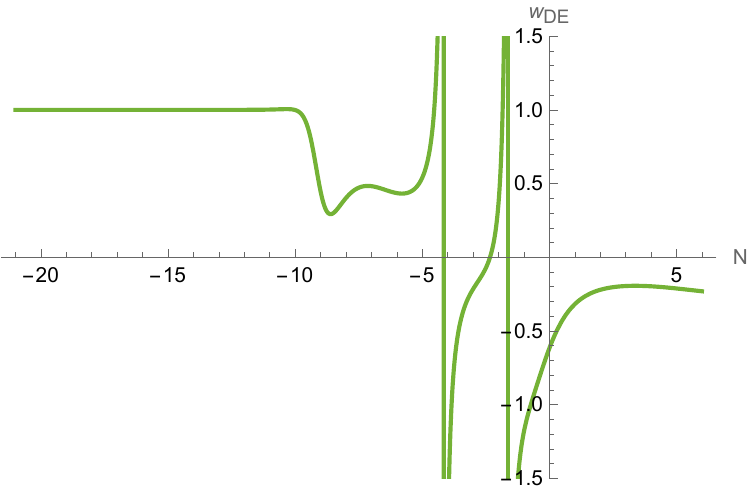}\caption{}\label{fig:ExpExpradwN}
\end{subfigure}\quad
\begin{subfigure}[H]{0.48\textwidth}
\includegraphics[width=\textwidth]{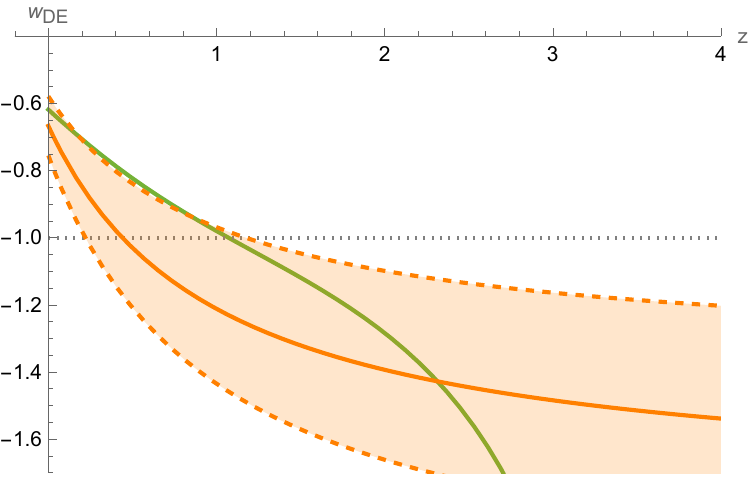}\caption{}\label{fig:ExpExpradwzobs}
\end{subfigure}
\caption{Cosmological solution in the string-inspired model with exponential potential and exponential couplings, as given in \eqref{solstring2}. The evolution of the $\Omega_n$ is provided in Figure \ref{fig:ExpExpradOps} for $\Omega_{r,m,\varphi}$, and Figure \ref{fig:ExpExpradOs} for $\bar{\Omega}_{r,m}, \Omega_{{\rm DE}}$. Figure \ref{fig:ExpExpradwN} and \ref{fig:ExpExpradwzobs} display the evolution of the effective dark energy equation of state parameter $w_{{\rm DE}}$ in green, the last Figure giving in addition the comparison to observational values as described for Figure \ref{fig:ExpExp3wzfit}.}\label{fig:ExpExprad}
\end{center}
\end{figure}

We now make use of the above values in slightly extended coupling functions
\beq
A_{m,r} = 1 + A_{m,r \, \varphi} (e^{\alpha_{m,r}\, \varphi} -1 ) \ ,
\eeq
that allow, as before, some part of the matter and radiation to be decoupled. We then consider the following string-inspired values for this second solution
\beq
\lambda=\sqrt{\frac{8}{3}}\ ,\ \alpha_m=\sqrt{\frac{3}{8}}\ ,\ \alpha_r=\sqrt{\frac{3}{2}}\ ,\  A_{m\,\varphi}=A_{r\,\varphi}= \frac{1}{8}\ ,\quad {\rm and}\ w_{\varphi0}=-0.61865 \ ,\label{solstring2}
\eeq
giving the solution displayed in Figure \ref{fig:ExpExprad}.
We see in Figure \ref{fig:ExpExpradOps} and \ref{fig:ExpExpradOs} a standard past history of the universe up to the point of radiation - matter equality; beyond this, the solution displays surprising features that would be worth investigating in more details. We discard this far past here, since we are mostly interested in the recent evolution of $w_{{\rm DE}}$. The latter displays a phantom regime as expected, with phantom crossing at $z_c = 1.07$. We also note from Figure \ref{fig:ExpExpradwzobs} a fairly good agreement to the observational constraints.\\

We conclude that steep stringy potentials, together with string-inspired coupling functions, provide an observationally competitive evolution of the dark energy equation of state parameter $w_{{\rm DE}}$, with in particular a phantom regime. For these cosmological solutions and others, we now turn to studying some interesting features.

\section{Appearance of poles in $w_{{\rm DE}}$, and observations}\label{sec:poles}

Looking at Figure \ref{fig:ExpExpradwN}, we see the presence of 2 divergences in $w_{{\rm DE}}$: as we will see, those are due to poles, that we discuss in this section. Interestingly, poles do not appear for the solution displayed in Figure \ref{fig:ExpExp3wN}: we will also study how varying solution parameters make them appear, and finally compare this situation to observations. In the following, we focus on the case $A_r=0$, i.e.~with only matter coupling for simplicity; the discussion could be extended to include a radiation coupling. We take this opportunity to present other cosmological solutions with a valid radiation domination phase in the past.

\subsection{Poles and their properties}

As was made clear in Section \ref{sec:form}, the way dark energy is defined is effective, and with respect to observations, but the related measured quantities $w_{{\rm DE}}$ and $\Omega_{{\rm DE}}$ are not really physical. We already explained that $-1 \leq w_{\varphi} \leq 1$, and $\rho_{\varphi}>0$ should remain finite, at least in the recent universe. Those physical quantities are to be contrasted with $w_{{\rm DE}}$ and $\rho_{{\rm DE}}$ which are not physical but effective. Given the definition $w_{{\rm DE}}= \frac{w_{\varphi} \rho_{\varphi}}{\rho_{{\rm DE}}} $ and the finiteness of the numerator, a divergence of $w_{{\rm DE}}$ appears at a pole
\beq
{\rm Pole:}\quad \rho_{{\rm DE}} =0 \ \Leftrightarrow \ w_{{\rm DE}} \rightarrow \pm \infty\ .
\eeq

\subsubsection*{Appearance of the first pole}

Can such a pole appear in a realistic cosmological solution? We recall the definition $\rho_{{\rm DE}} = \rho_{\varphi} + (A_m-1) \bar{\rho}_m $. Let us first note that $\rho_{\varphi}$ and $\bar{\rho}_m$ are both positive, and in the recent past at least, $A_m-1<0$: it then amounts to a competition between these terms as to whether their sum would vanish. In addition, one reads from continuity equations of $\rho_{\varphi}$ and $\bar{\rho}_m$ that they both decrease with time (ensuring in particular them to stay positive in the past). Whether $\rho_{{\rm DE}}$ would vanish in the past is then not obvious and strongly depends on the coupling function $A_m$.

Another way to see the possibility of the appearance of a first pole is as follows. As pointed-out previously, one can read from the continuity equation of $\rho_{{\rm DE}}$ that it reaches a maximum at phantom crossing $z_c$. Before $z_c$, one thus has $\rho_{{\rm DE}}$ growing with time. In addition, between $z_c$ and today, $\rho_{{\rm DE}}$ has to be positive. Going backwards in time, $\rho_{{\rm DE}}$ then decreases towards $0$ and it is possible that it decreases enough to vanish. As mentioned, hitting this pole or not depends on the details of $A_m$. We will investigate below this dependence.

\subsubsection*{Around the first pole}

If a first pole is reached, it is important to note that the product $w_{{\rm DE}} \rho_{{\rm DE}} = w_{\varphi} \rho_{\varphi}$ remains finite and continuous. As a consequence, we see from the continuity equation of $\rho_{{\rm DE}}$ that its derivative undergoes no discontinuity, and is thus positive all around the pole. We deduce that $\rho_{{\rm DE}}<0$ just before the pole! This change of sign of $\rho_{{\rm DE}}$ also implies a change of sign of $w_{{\rm DE}}$, i.e.~$w_{{\rm DE}}>0$ just before the pole (see \cite{Ozulker:2022slu} for related discussions).

Turning to the $\Omega_n$, interesting observations can be made. The pole corresponds to $\Omega_{{\rm DE}}=0$. Given the first Friedmann equation, and neglecting radiation at this first pole, one obtains
\beq
\text{First pole:}\quad \Omega_{{\rm DE}} =0 \ \Leftrightarrow \ \bar{\Omega}_m \approx 1\ .
\eeq
Note that this corresponds, in terms of the other variables, to $\Omega_m\approx A_m \approx 1 -\Omega_{\varphi}$ at this first pole. Following up on the above, we deduce that
\beq
\text{Just before first pole:}\quad \Omega_{{\rm DE}} <0 \ \Leftrightarrow \ \bar{\Omega}_m > 1\ ,
\eeq
which is certainly unusual with respect to the standard $\Omega_n$.\\

\begin{figure}[H]
\begin{center}
\begin{subfigure}[H]{0.48\textwidth}
\includegraphics[width=\textwidth]{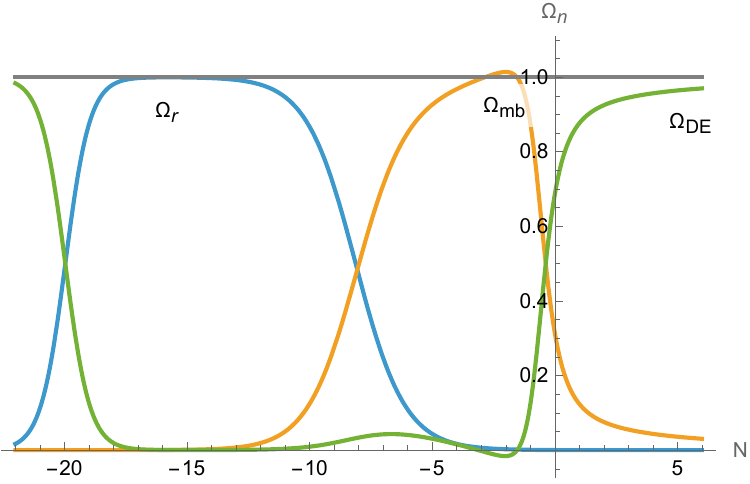}\caption{}\label{fig:ExpExp1Os}
\end{subfigure}\quad
\begin{subfigure}[H]{0.48\textwidth}
\includegraphics[width=\textwidth]{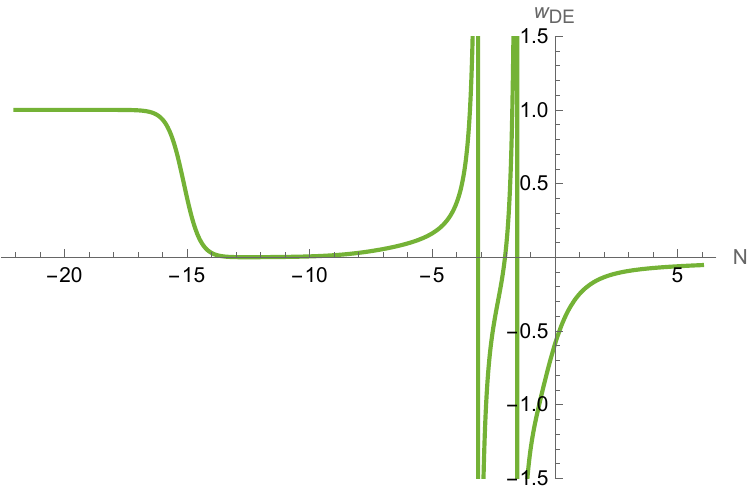}\caption{}\label{fig:ExpExp1wN}
\end{subfigure}
\caption{Cosmological solution for exponential potential and exponential coupling, as given in \eqref{solExpExp1}. The evolution of $\Omega_{r}, \bar{\Omega}_{m}, \Omega_{{\rm DE}}$ is provided in Figure \ref{fig:ExpExp1Os}, together with a gray line at $\Omega_n=1$. Figure \ref{fig:ExpExp1wN} displays the evolution of the effective dark energy equation of state parameter $w_{{\rm DE}}$.}\label{fig:ExpExp1}
\end{center}
\end{figure}

Let us illustrate the properties discussed so far. We consider the phenomenological model\footnote{This model is very close to a linear coupling with $A_m(\varphi) = 1+ \frac{1}{10} ( \varphi  - \varphi_0 )$ and $\lambda=\sqrt{3}$. For the latter we obtain $w_{\varphi0}=-0.59117876830$ and almost identical graphs compared to Figure \ref{fig:ExpExp1}.}
\beq
\lambda=\sqrt{3}\ ,\ A_m(\varphi) = 2 - e^{-\frac{1}{10} \, \varphi} \ ,\ w_{\varphi0}=-0.59388487348 \ ,\label{solExpExp1}
\eeq
giving the solution in Figure \ref{fig:ExpExp1}.
We observe in Figure \ref{fig:ExpExp1} the properties analytically described before. The first pole in $w_{{\rm DE}}$ appears around $N\approx -1.5$. We verify that $w_{{\rm DE}}$ is positive earlier and negative later. Turning to $\Omega_n$, this first pole corresponds to $\Omega_{{\rm DE}} =0 \ ,\ \bar{\Omega}_m \approx 1$, and we verify that earlier, $\Omega_{{\rm DE}} <0 \ ,\ \bar{\Omega}_m > 1$.

\subsubsection*{Second pole}

\begin{figure}[H]
\begin{center}
\begin{subfigure}[H]{0.48\textwidth}
\includegraphics[width=\textwidth]{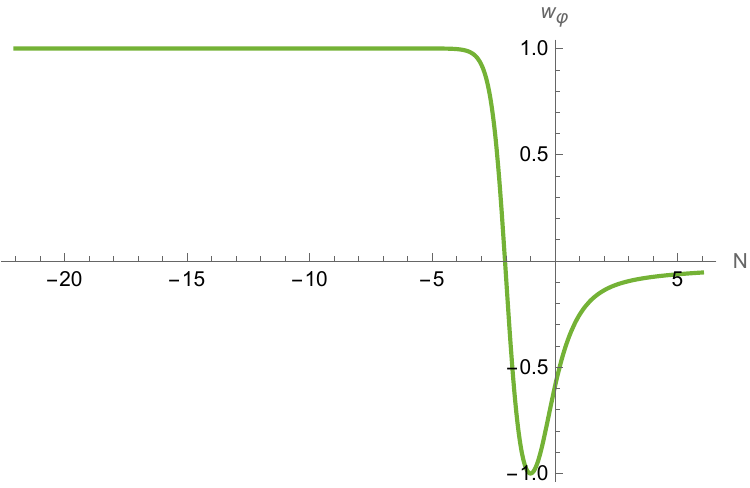}\caption{}\label{fig:ExpExp1wpN}
\end{subfigure}\quad
\begin{subfigure}[H]{0.48\textwidth}
\includegraphics[width=\textwidth]{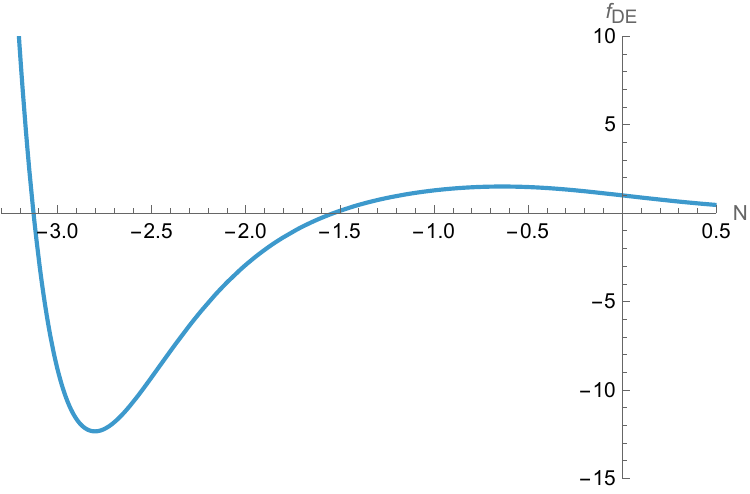}\caption{}\label{fig:ExpExp1fDEN}
\end{subfigure}
\caption{Evolution of $w_{\varphi}(N)$ and $f_{{\rm DE}}(N)$ in the cosmological solution with exponential potential and exponential coupling, as given in \eqref{solExpExp1} and Figure \ref{fig:ExpExp1}. The profile of $\varphi(N)$, not displayed here, is very similar to the one in Figure \ref{fig:ExpExp3phiN}.}\label{fig:ExpExp1wf}
\end{center}
\end{figure}

We observe at $N\approx -3$ on Figure \ref{fig:ExpExp1} (as well as on Figure \ref{fig:ExpExpradwN}) the presence of a second pole in $w_{{\rm DE}}$ at earlier times. Without surprise, this matches the moment when $\Omega_{{\rm DE}} =0$. As before, $w_{{\rm DE}} \rho_{{\rm DE}} = w_{\varphi} \rho_{\varphi} $ is finite and continuous. As can be inferred from the sign of the left-hand side, later than the second pole ($\rho_{{\rm DE}} <0$, $w_{{\rm DE}} <0$), we have $w_{\varphi}>0$; we also verify this in Figure \ref{fig:ExpExp1wpN}. Earlier than the second pole, $\rho_{{\rm DE}} >0$, so we deduce that $w_{{\rm DE}} >0$. This is confirmed in Figure \ref{fig:ExpExp1wN}.

The reasons for this evolution and second pole are less clear, but seem related to the growth of the radiation contribution $\Omega_r$, backwards in time: radiation grows, while matter diminishes after having reached its maximum. The latter coincides approximately with a negative minimum of $\Omega_{{\rm DE}}$, which then goes back to vanishing. We have
\beq
\text{Second pole:}\quad \Omega_{{\rm DE}} =0 \ \Leftrightarrow \ \bar{\Omega}_m = 1 - \Omega_r\ .
\eeq

We finally provide in Figure \ref{fig:ExpExp1wf} further elements of this solution example, to help the understanding.

\subsection{Appearance of poles and model parameters}

\begin{figure}[H]
\begin{center}
\begin{subfigure}[H]{0.48\textwidth}
\includegraphics[width=\textwidth]{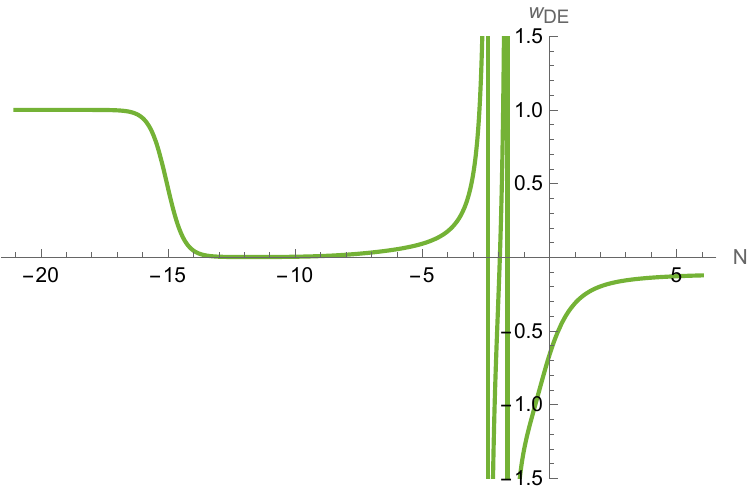}\caption{$\lambda=\sqrt{\frac{8}{3}}, \ A_{m\varphi}= \frac{1}{8},\ w_{\varphi0}=-0.67,\ z_c =0.73$}\label{fig:ExpExp2wN}
\end{subfigure}\quad
\begin{subfigure}[H]{0.48\textwidth}
\includegraphics[width=\textwidth]{figExpExp5wN.pdf}\caption{$\lambda=\sqrt{2}, \ A_{m\varphi}= \frac{1}{16},\ w_{\varphi0}=-0.73, \ z_c=0.93$}\label{fig:ExpExp5wN}
\end{subfigure}\\
\begin{subfigure}[H]{0.48\textwidth}
\includegraphics[width=\textwidth]{figExpExp3wN.pdf}\caption{$\lambda=\sqrt{2}, \ A_{m\varphi}= \frac{1}{8},\ w_{\varphi0}=-0.76,\ z_c =0.64$}\label{fig:ExpExp3wNbis}
\end{subfigure}\quad
\begin{subfigure}[H]{0.48\textwidth}
\includegraphics[width=\textwidth]{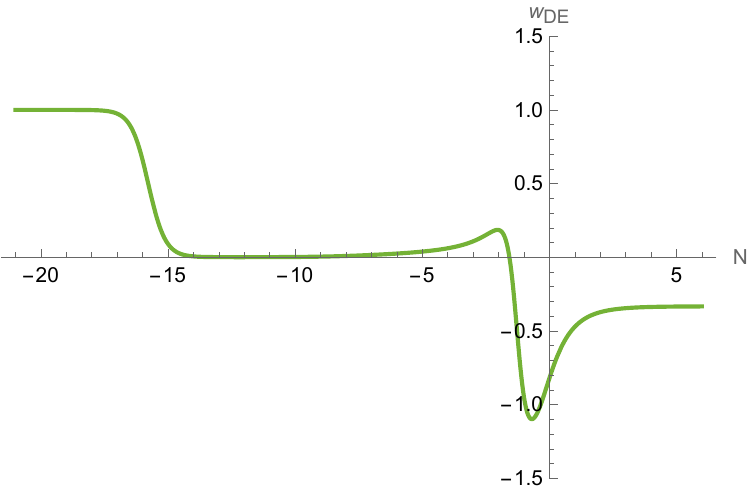}\caption{$\lambda=\sqrt{2}, \ A_{m\varphi}= \frac{1}{4},\ w_{\varphi0}=-0.82,\ z_c=0.40$}\label{fig:ExpExp4wN}
\end{subfigure}
\caption{Evolution of $w_{{\rm DE}}(N)$ along cosmological solutions for various model parameters.}\label{fig:ExpExpwNall}
\end{center}
\end{figure}

We now make the model parameters vary, considering a matter coupling of the form
\beq
A_m(\varphi) = 1+A_{m\varphi} (1 - e^{-\alpha_m \varphi}) \ ,\ \alpha_m=\sqrt{\frac{2}{3}} \ ,
\eeq
and an exponential potential. We find cosmological solutions with past radiation domination in a small range of parameters. We display our examples in Figure \ref{fig:ExpExpwNall}.
As we can see in Figure \ref{fig:ExpExpwNall}, the appearance of poles is very sensitive to the parameter values. In the recent work \cite{Chakraborty:2025syu}, that also considers exponential potential and coupling, poles seem to appear, although not explicitly shown. We see there as well important variations of $w_{{\rm DE}}(N)$ with small changes of model parameters. For us here, the poles appear for higher $\lambda$ or lower $A_{m\varphi}$, the range over which this changes being however very small. As mentioned previously, we certainly do not know about detailed factors of 2 in $A_m$ at this stage, but the appearance of poles depends on this. This provides motivation to work-out in detail a string theory model with matter coupling.

\subsection{Poles and observations}

Whether $w_{{\rm DE}}$ is depicted in terms of $N,a$ or $z$, we note the following important point: if no pole appears, $w_{{\rm DE}}$ reaches a finite minimum in the recent universe. Around the latter, the function is then {\sl convex}. On the contrary, if a pole appears, the divergence makes the function being {\sl concave} close to the pole. This can be seen in the examples of Figure \ref{fig:ExpExpwNall} for $w_{{\rm DE}}(N)$, or in the observationally compatible examples of Figure \ref{fig:ExpExp3wz} and \ref{fig:ExpExpradwzobs} for $w_{{\rm DE}}(z)$. As can be seen in these last two examples, this distinction between convex and concave, that could inform us about a pole, could be detected already by DESI before $z\lesssim 4$. Equivalently, a pole would appear if $f_{{\rm DE}}$ or $\Omega_{{\rm DE}}$ vanishes, which should also be observable by DESI in the recent universe.

Regarding $w_{{\rm DE}}$, unfortunately, the CPL parametrisation does not inform us on this matter: $w_{{\rm DE}}(a)$ is precisely in-between concave and convex, as a linear function. Figure \ref{fig:ExpExp3wafit} illustrates how a minimum can then be missed. The corresponding $w_{{\rm DE}}(z)$ obtained from CPL is systematically convex without ever reaching a minimum. To get a hint on this question, we can look at the reconstructions of these functions that have been carried out for DESI DR1 \cite{DESI:2024aqx} and DR2 \cite{DESI:2025fii}. The comparison of the latter two can be found in Fig. 17 and 18 of \cite{DESI:2025fii}. Those essentially use DESI + CMB + Union3 as data sets.

Two reconstructions methods are proposed, which however give different answers; this situation is reminiscent of the noticed high sensitivity to the model parameters. A first reconstruction method uses Chebyshev polynomials. This gives around $z=2.5$ a $w_{{\rm DE}}(z)$ which seems concave: see e.g.~\cite[Fig.5]{DESI:2025fii}. Another method uses Gaussian Process (GP) regression. This gives around $z=3$ a seemingly convex $w_{{\rm DE}}(z)$: see e.g.~\cite[Fig.9,10]{DESI:2025fii}. In both cases, $f_{{\rm DE}}(z)$ does not get negative (see however results without CMB in \cite{DESI:2024aqx}), but can go lower than $0.5$ in the range of $z$ considered.

This observational situation calls for a more thorough analysis, in particular extending it to $z=4$. This could provide important information on the details of the model to be used (giving poles or not), as motivated by the above discussion.

\section{Outlook: Early Dark Energy for free?}\label{sec:out}

Motivated by the reported observations of a phantom dark energy in the recent universe, we have considered in this work quintessence models with a coupling to matter (and possibly radiation). Those are known to reproduce a phantom regime for an effective dark energy, as reviewed in Section \ref{sec:form}. We have explained, and illustrated in Section \ref{sec:string}, that {\sl steep} potentials together with a matter coupling could provide realistic cosmological solutions in very good agreement with observations. It is of particular interest for asymptotic exponential potentials of string theory, whose exponential rate obeys $\lambda \geq \sqrt{2}$, since those are otherwise disfavored by observations without matter-coupling. This motivates for future work the precise determination of matter (and radiation) couplings in string models of particle physics. We finally discussed in Section \ref{sec:poles} the appearance of poles in $w_{{\rm DE}}(z)$ for certain choices of model parameters, for $z\lesssim 4$. Observations could foresee such poles, or their absence, by checking whether the function $w_{{\rm DE}}(z)$ gets concave or convex in the past, providing this way valuable information on an adequate model.\\

\begin{figure}[t]
\begin{center}
\begin{subfigure}[H]{0.48\textwidth}
\includegraphics[width=\textwidth]{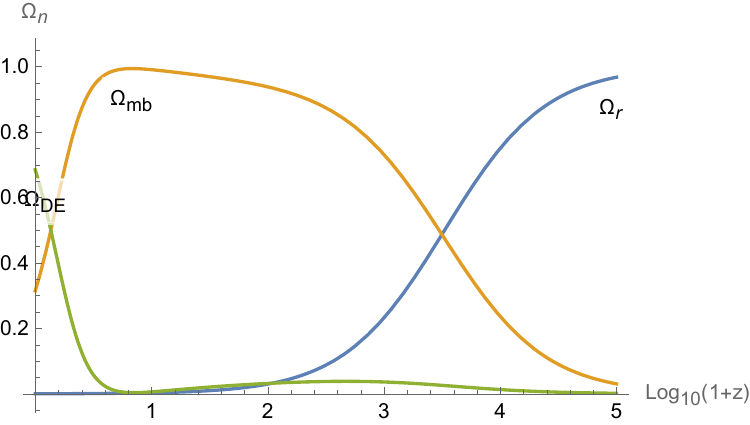}\caption{}\label{fig:ExpExp3EDErho}
\end{subfigure}\quad
\begin{subfigure}[H]{0.48\textwidth}
\includegraphics[width=\textwidth]{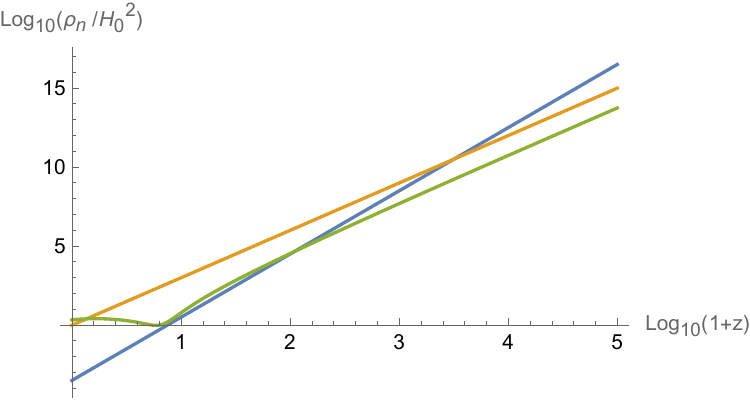}\caption{}\label{fig:ExpExp3EDEOz}
\end{subfigure}
\caption{Evolution of $\Omega_r, \bar{\Omega}_m, \Omega_{{\rm DE}}$, and the corresponding $\log_{10}(\rho_n / H_0^2)$, in terms of $\log_{10}(1+z)$, for the cosmological solution of Figure \ref{fig:ExpExp3}. One can notice a bump in $\Omega_{{\rm DE}}$ around $z \approx 10^3$, very reminiscent of Early Dark Energy models, used to solve the Hubble tension. This feature comes for free and systematically out of the various the models considered in this work.}\label{fig:ExpExp3EDE}
\end{center}
\end{figure}

We end with a comment on realistic cosmological solutions obtained throughout this paper. When considering the effectively observed quantities $\Omega_r, \bar{\Omega}_m, \Omega_{{\rm DE}}$, we notice systematically a little bump or maximum in the dark energy constituent $\Omega_{{\rm DE}}$, approximately around $z \sim 10^3$. We display this in various fashions in Figure \ref{fig:ExpExp3EDE} for the first string-inspired example of Section \ref{sec:string} (Figure \ref{fig:ExpExp3}, \ref{fig:ExpExp3wp} and \ref{fig:ExpExp3fit}), but the same feature was noticed in the other models and solutions. This behaviour is very reminiscent of Early Dark Energy (EDE) models \cite{Kamionkowski:2022pkx, Poulin:2023lkg}, which help to solve the Hubble tension (see however \cite{Vagnozzi:2023nrq}).

Interestingly, this feature comes for free, as an output from these models. A closer look at
\beq
\Omega_{{\rm DE}} = \Omega_{\varphi} + (A_m -1) \bar{\Omega}_m \ ,
\eeq
in the case without radiation-coupling, indicates that this EDE-like bump is due to the second term, dominant over the first one. Indeed, the field evolution (see Figure \ref{fig:ExpExp3phiN}) is such that $A_m -1$ becomes positive and non-negligible in the relevant redshift range, while $\bar{\Omega}_m$ is then dominant. Such an evolution is systematic in the cosmological solutions considered from these models, hence the systematic feature.

Can this really help to solve the Hubble tension?\footnote{We note that $w_{{\rm DE}}\geq0$ during this bump; it is unclear to us how adequate that is.} The precise values of $z$ and of the amplitude of this dark energy bump are highly sensitive to the model details. We therefore refrain from providing an evaluation of this effect. One would also need to determine precisely when to place the CMB in this slightly modified universe history. It remains appealing that such a feature comes out for free and systematically, from the models considered so far to reproduce the phantom regime while accommodating asymptotic stringy potentials. More than ever, it seems that Phantom matters.

\subsection*{Acknowledgements}

We thank P.~Brax, C.~Delaunay, L.~Pinol, D.~Shlivko, J.P.~Uzan, T.~Wrase and C.~Yeche for helpful discussions during the completion of this work. We also thank the organisers of the workshop {\it Cosmology Beyond $\Lambda$CDM}, held in PCTS, Princeton, in May 2025, for a fruitful and inspiring event.

\begin{appendix}

\section{Hilltop potential and linear coupling}\label{ap:hilltop}

Realistic cosmological solutions are not easy to find within the models considered; observational features of the solutions are highly dependent on the model parameters, which thus require tuning. In this appendix, we present a different model, without exponentials: we consider a hilltop potential and a linear matter-coupling
\beq
V(\varphi) = V_0 \left( 1- \frac{\kappa^2}{2} \varphi^2 \right) \ ,\quad A_m(\varphi)= 1 + A_{m\varphi} (\varphi - \varphi_0) \ .
\eeq
The hilltop potential can be viewed as part of an axionic potential (with cosine), so a linear (axion-like) coupling seems adequate. Such a linear coupling is also reminiscent of a Yukawa coupling. We also note that this provides a growing function $A_m(\varphi)$, as desired.

\begin{figure}[t]
\begin{center}
\begin{subfigure}[H]{0.48\textwidth}
\includegraphics[width=\textwidth]{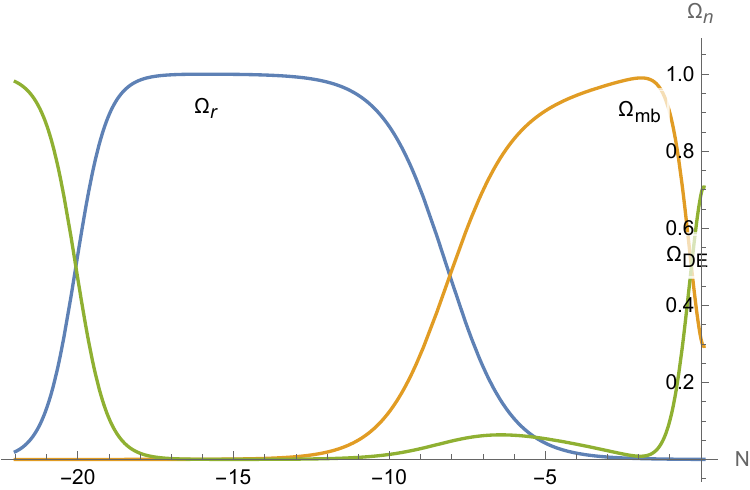}\caption{}\label{fig:HillLinOs}
\end{subfigure}\quad
\begin{subfigure}[H]{0.48\textwidth}
\includegraphics[width=\textwidth]{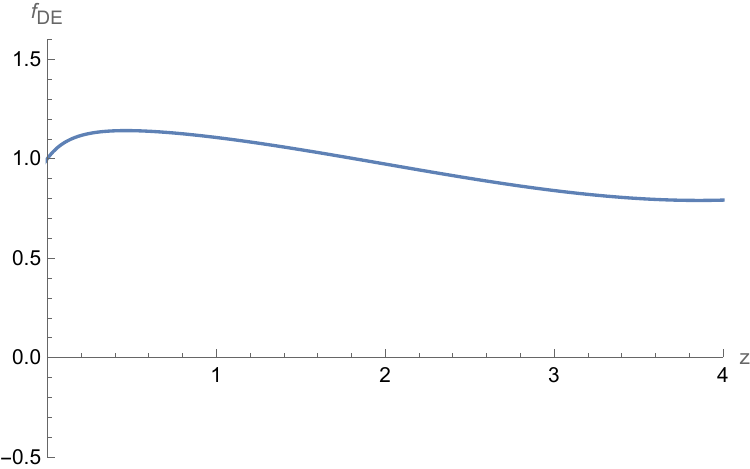}\caption{}\label{fig:HillLinfDEz}
\end{subfigure}\\
\begin{subfigure}[H]{0.48\textwidth}
\includegraphics[width=\textwidth]{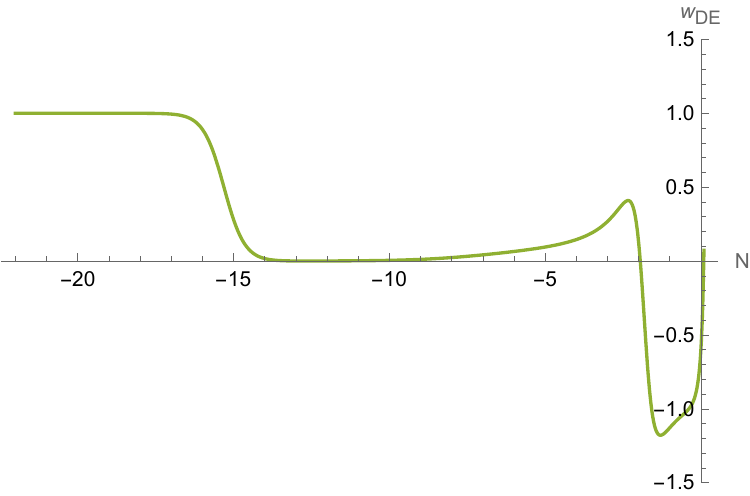}\caption{}\label{fig:HillLinwN}
\end{subfigure}\quad
\begin{subfigure}[H]{0.48\textwidth}
\includegraphics[width=\textwidth]{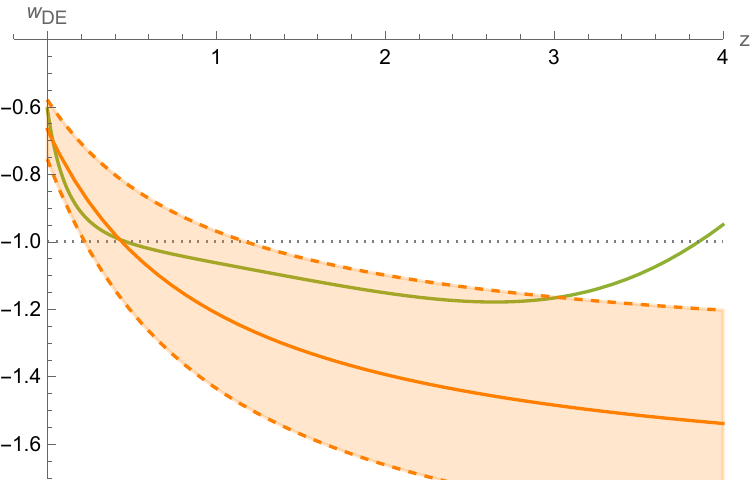}\caption{}\label{fig:HillLinwzobs}
\end{subfigure}
\caption{Cosmological solution in a model with hilltop potential and linear matter-coupling, as detailed in the main text. The graphs are analogous to those of Figure \ref{fig:ExpExp3} and \ref{fig:ExpExp3fit}, to which we refer for more details. We see that $w_{{\rm DE}}$ does not admit poles. The solution is in agreement with observations only until $z \approx 3$.}\label{fig:HillLin}
\end{center}
\end{figure}

To obtain numerically a cosmological solution, we follow the procedure described in Section \ref{sec:string} and in \cite{Andriot:2024sif}. With this potential, we now fix the initial condition
\beq
\varphi_0 = \sqrt{\frac{2}{\kappa^2}\left(1 - \frac32 \frac{\Omega_{\varphi0}}{\tilde{V}_0}(1 - w_{\varphi0})\right)} \ ,
\eeq
and fix the potential parameters to chosen values. Here, we pick
\beq
\tilde{V}_0=\frac{5}{2} \ ,\ \kappa = 3 \ , \ A_{m\varphi}=\frac{1}{10} \ ,\ {\rm and}\ w_{\varphi0}=-0.60618663271 \ ,
\eeq
which gives the cosmological solution depicted in Figure \ref{fig:HillLin}.

While qualitatively comparable to other solutions obtained in this paper, we see that the solution of Figure \ref{fig:HillLin} is only in agreement with observations for low $z$. We have difficulties to do better with hilltop potential and linear coupling. Exponentials, as in Figure \ref{fig:ExpExp3}, may then be more promising, contrary to the situation without coupling to matter \cite{Shlivko:2024llw}.

\end{appendix}

\newpage

\addcontentsline{toc}{section}{References}

\providecommand{\href}[2]{#2}\begingroup\raggedright\endgroup


\begin{thebibliography}{10}

\bibitem{DES:2024jxu}
{\scshape DES} collaboration, T.~M.~C. Abbott et~al., \emph{{The Dark Energy
  Survey: Cosmology Results with \ensuremath{\sim}1500 New High-redshift Type
  Ia Supernovae Using the Full 5 yr Data Set}},
  \href{http://dx.doi.org/10.3847/2041-8213/ad6f9f}{\emph{Astrophys. J. Lett.}
  {\bfseries 973} (2024) L14},
  [\href{https://arxiv.org/abs/2401.02929}{{\ttfamily 2401.02929}}].

\bibitem{DESI:2024mwx}
{\scshape DESI} collaboration, A.~G. Adame et~al., \emph{{DESI 2024 VI:
  cosmological constraints from the measurements of baryon acoustic
  oscillations}},
  \href{http://dx.doi.org/10.1088/1475-7516/2025/02/021}{\emph{JCAP} {\bfseries
  02} (2025) 021}, [\href{https://arxiv.org/abs/2404.03002}{{\ttfamily
  2404.03002}}].

\bibitem{DESI:2025zgx}
{\scshape DESI} collaboration, M.~Abdul~Karim et~al., \emph{{DESI DR2 Results
  II: Measurements of Baryon Acoustic Oscillations and Cosmological
  Constraints}},  [\href{https://arxiv.org/abs/2503.14738}{{\ttfamily
  2503.14738}}].

\bibitem{Ratra:1987rm}
B.~Ratra and P.~J.~E. Peebles, \emph{{Cosmological Consequences of a Rolling
  Homogeneous Scalar Field}},
  \href{http://dx.doi.org/10.1103/PhysRevD.37.3406}{\emph{Phys. Rev. D}
  {\bfseries 37} (1988) 3406}.

\bibitem{Peebles:1987ek}
P.~J.~E. Peebles and B.~Ratra, \emph{{Cosmology with a Time Variable
  Cosmological Constant}},
  \href{http://dx.doi.org/10.1086/185100}{\emph{Astrophys. J. Lett.} {\bfseries
  325} (1988) L17}.

\bibitem{Wetterich:1994bg}
C.~Wetterich, \emph{{The Cosmon model for an asymptotically vanishing time
  dependent cosmological 'constant'}}, {\emph{Astron. Astrophys.} {\bfseries
  301} (1995) 321--328},
  [\href{https://arxiv.org/abs/hep-th/9408025}{{\ttfamily hep-th/9408025}}].

\bibitem{Caldwell:1997ii}
R.~R. Caldwell, R.~Dave and P.~J. Steinhardt, \emph{{Cosmological imprint of an
  energy component with general equation of state}},
  \href{http://dx.doi.org/10.1103/PhysRevLett.80.1582}{\emph{Phys. Rev. Lett.}
  {\bfseries 80} (1998) 1582--1585},
  [\href{https://arxiv.org/abs/astro-ph/9708069}{{\ttfamily
  astro-ph/9708069}}].

\bibitem{Andriot:2024sif}
D.~Andriot, \emph{{Quintessence: an analytical study, with theoretical and
  observational applications}},
  [\href{https://arxiv.org/abs/2410.17182}{{\ttfamily 2410.17182}}].

\bibitem{DESI:2025fii}
{\scshape DESI} collaboration, K.~Lodha et~al., \emph{{Extended Dark Energy
  analysis using DESI DR2 BAO measurements}},
  [\href{https://arxiv.org/abs/2503.14743}{{\ttfamily 2503.14743}}].

\bibitem{Anchordoqui:2025fgz}
L.~A. Anchordoqui, I.~Antoniadis and D.~Lust, \emph{{S-dual Quintessence, the
  Swampland, and the DESI DR2 Results}},
  [\href{https://arxiv.org/abs/2503.19428}{{\ttfamily 2503.19428}}].

\bibitem{Rudelius:2021oaz}
T.~Rudelius, \emph{{Dimensional reduction and (Anti) de Sitter bounds}},
  \href{http://dx.doi.org/10.1007/JHEP08(2021)041}{\emph{JHEP} {\bfseries 08}
  (2021) 041}, [\href{https://arxiv.org/abs/2101.11617}{{\ttfamily
  2101.11617}}].

\bibitem{Bedroya:2019snp}
A.~Bedroya and C.~Vafa, \emph{{Trans-Planckian Censorship and the Swampland}},
  \href{http://dx.doi.org/10.1007/JHEP09(2020)123}{\emph{JHEP} {\bfseries 09}
  (2020) 123}, [\href{https://arxiv.org/abs/1909.11063}{{\ttfamily
  1909.11063}}].

\bibitem{Agrawal:2018own}
P.~Agrawal, G.~Obied, P.~J. Steinhardt and C.~Vafa, \emph{{On the Cosmological
  Implications of the String Swampland}},
  \href{http://dx.doi.org/10.1016/j.physletb.2018.07.040}{\emph{Phys. Lett. B}
  {\bfseries 784} (2018) 271--276},
  [\href{https://arxiv.org/abs/1806.09718}{{\ttfamily 1806.09718}}].

\bibitem{Akrami:2018ylq}
Y.~Akrami, R.~Kallosh, A.~Linde and V.~Vardanyan, \emph{{The Landscape, the
  Swampland and the Era of Precision Cosmology}},
  \href{http://dx.doi.org/10.1002/prop.201800075}{\emph{Fortsch. Phys.}
  {\bfseries 67} (2019) 1800075},
  [\href{https://arxiv.org/abs/1808.09440}{{\ttfamily 1808.09440}}].

\bibitem{Raveri:2018ddi}
M.~Raveri, W.~Hu and S.~Sethi, \emph{{Swampland Conjectures and Late-Time
  Cosmology}}, \href{http://dx.doi.org/10.1103/PhysRevD.99.083518}{\emph{Phys.
  Rev. D} {\bfseries 99} (2019) 083518},
  [\href{https://arxiv.org/abs/1812.10448}{{\ttfamily 1812.10448}}].

\bibitem{Schoneberg:2023lun}
N.~Sch\"oneberg, L.~Vacher, J.~D.~F. Dias, M.~M. C.~D. Carvalho and C.~J. A.~P.
  Martins, \emph{{News from the Swampland \textemdash{} constraining string
  theory with astrophysics and cosmology}},
  \href{http://dx.doi.org/10.1088/1475-7516/2023/10/039}{\emph{JCAP} {\bfseries
  10} (2023) 039}, [\href{https://arxiv.org/abs/2307.15060}{{\ttfamily
  2307.15060}}].

\bibitem{Bhattacharya:2024hep}
S.~Bhattacharya, G.~Borghetto, A.~Malhotra, S.~Parameswaran, G.~Tasinato and
  I.~Zavala, \emph{{Cosmological constraints on curved quintessence}},
  \href{http://dx.doi.org/10.1088/1475-7516/2024/09/073}{\emph{JCAP} {\bfseries
  09} (2024) 073}, [\href{https://arxiv.org/abs/2405.17396}{{\ttfamily
  2405.17396}}].

\bibitem{Alestas:2024gxe}
G.~Alestas, M.~Delgado, I.~Ruiz, Y.~Akrami, M.~Montero and S.~Nesseris,
  \emph{{Is curvature-assisted quintessence observationally viable?}},
  \href{http://dx.doi.org/10.1103/PhysRevD.110.106010}{\emph{Phys. Rev. D}
  {\bfseries 110} (2024) 106010},
  [\href{https://arxiv.org/abs/2406.09212}{{\ttfamily 2406.09212}}].

\bibitem{Akrami:2025zlb}
Y.~Akrami, G.~Alestas and S.~Nesseris, \emph{{Has DESI detected exponential
  quintessence?}},  [\href{https://arxiv.org/abs/2504.04226}{{\ttfamily
  2504.04226}}].

\bibitem{Planck:2018vyg}
{\scshape Planck} collaboration, N.~Aghanim et~al., \emph{{Planck 2018 results.
  VI. Cosmological parameters}},
  \href{http://dx.doi.org/10.1051/0004-6361/201833910}{\emph{Astron.
  Astrophys.} {\bfseries 641} (2020) A6},
  [\href{https://arxiv.org/abs/1807.06209}{{\ttfamily 1807.06209}}].

\bibitem{Ludwick:2017tox}
K.~J. Ludwick, \emph{{The viability of phantom dark energy: A review}},
  \href{http://dx.doi.org/10.1142/S0217732317300257}{\emph{Mod. Phys. Lett. A}
  {\bfseries 32} (2017) 1730025},
  [\href{https://arxiv.org/abs/1708.06981}{{\ttfamily 1708.06981}}].

\bibitem{Escamilla:2023oce}
L.~A.~Escamilla, W.~Giar\`e, E.~Di Valentino, R.~C.~Nunes and S.~Vagnozzi, \emph{The state of the dark energy equation of state circa 2023}, \emph{JCAP} \textbf{05} (2024) 091, [\href{https://arxiv.org/abs/2307.14802}{{\ttfamily 2307.14802}}].

\bibitem{ACT:2025tim}
{\scshape ACT} collaboration, E.~Calabrese et~al., \emph{{The Atacama Cosmology
  Telescope: DR6 Constraints on Extended Cosmological Models}},
  [\href{https://arxiv.org/abs/2503.14454}{{\ttfamily 2503.14454}}].

\bibitem{Nesseris:2025lke}
S.~Nesseris, Y.~Akrami and G.~D. Starkman, \emph{{To CPL, or not to CPL? What
  we have not learned about the dark energy equation of state}},
  [\href{https://arxiv.org/abs/2503.22529}{{\ttfamily 2503.22529}}].

\bibitem{Shlivko:2025fgv}
D.~Shlivko, P.~J. Steinhardt and C.~L. Steinhardt, \emph{{Optimal
  parameterizations for observational constraints on thawing dark energy}},
  [\href{https://arxiv.org/abs/2504.02028}{{\ttfamily 2504.02028}}].

\bibitem{Li:2025cxn}
C.~Li, J.~Wang, D.~Zhang, E.~N. Saridakis and Y.-F. Cai, \emph{{Quantum Gravity
  Meets DESI: Dynamical Dark Energy in Light of the Trans-Planckian Censorship
  Conjecture}},  [\href{https://arxiv.org/abs/2504.07791}{{\ttfamily
  2504.07791}}].

\bibitem{Wu:2025wyk}
P.~J.~Wu, \emph{Comparison of dark energy models using late-universe observations}, [\href{https://arxiv.org/abs/2504.09054}{{\ttfamily
  2504.09054}}].

\bibitem{Caldwell:1999ew}
R.~R. Caldwell, \emph{{A Phantom menace?}},
  \href{http://dx.doi.org/10.1016/S0370-2693(02)02589-3}{\emph{Phys. Lett. B}
  {\bfseries 545} (2002) 23--29},
  [\href{https://arxiv.org/abs/astro-ph/9908168}{{\ttfamily
  astro-ph/9908168}}].

\bibitem{Tiwari:2024gzo}
Y.~Tiwari, U.~Upadhyay and R.~K. Jain, \emph{{Exploring cosmological imprints
  of phantom crossing with dynamical dark energy in Horndeski gravity}},
  \href{http://dx.doi.org/10.1103/PhysRevD.111.043530}{\emph{Phys. Rev. D}
  {\bfseries 111} (2025) 043530},
  [\href{https://arxiv.org/abs/2412.00931}{{\ttfamily 2412.00931}}].

\bibitem{Oriti:2025lwx}
D.~Oriti and X.~Pang, \emph{{Late-time cosmic acceleration from quantum
  gravity}},  [\href{https://arxiv.org/abs/2502.12419}{{\ttfamily 2502.12419}}].

\bibitem{Akarsu:2025gwi}
\"O.~Akarsu, L.~Perivolaropoulos, A.~Tsikoundoura, A.~E.~Y\"ukselci and A.~Zhuk, \emph{Dynamical dark energy with AdS-to-dS and dS-to-dS transitions: Implications for the $H_0$ tension}, [\href{https://arxiv.org/abs/2502.14667}{{\ttfamily 2502.14667}}].

\bibitem{Moghtaderi:2025cns}
E.~Moghtaderi, B.~R. Hull, J.~Quintin and G.~Geshnizjani, \emph{{How Much NEC
  Breaking Can the Universe Endure?}},
  [\href{https://arxiv.org/abs/2503.19955}{{\ttfamily 2503.19955}}].

\bibitem{Ye:2025ulq}
G.~Ye and Y.~Cai, \emph{{NEC violation and ''beyond Horndeski'' physics in
  light of DESI DR2}},  [\href{https://arxiv.org/abs/2503.22515}{{\ttfamily
  2503.22515}}].

\bibitem{Smirnov:2025yru}
J.~Smirnov, \emph{{Dynamical Dark Energy Emerges from Massive Gravity}},
  [\href{https://arxiv.org/abs/2505.03870}{{\ttfamily 2505.03870}}].

\bibitem{Amendola:1999qq}
L.~Amendola, \emph{{Scaling solutions in general nonminimal coupling
  theories}}, \href{http://dx.doi.org/10.1103/PhysRevD.60.043501}{\emph{Phys.
  Rev. D} {\bfseries 60} (1999) 043501},
  [\href{https://arxiv.org/abs/astro-ph/9904120}{{\ttfamily
  astro-ph/9904120}}].

\bibitem{Billyard:2000bh}
A.~P. Billyard and A.~A. Coley, \emph{{Interactions in scalar field
  cosmology}}, \href{http://dx.doi.org/10.1103/PhysRevD.61.083503}{\emph{Phys.
  Rev. D} {\bfseries 61} (2000) 083503},
  [\href{https://arxiv.org/abs/astro-ph/9908224}{{\ttfamily
  astro-ph/9908224}}].

\bibitem{Huey:2004qv}
G.~Huey and B.~D. Wandelt, \emph{{Interacting quintessence. The Coincidence
  problem and cosmic acceleration}},
  \href{http://dx.doi.org/10.1103/PhysRevD.74.023519}{\emph{Phys. Rev. D}
  {\bfseries 74} (2006) 023519},
  [\href{https://arxiv.org/abs/astro-ph/0407196}{{\ttfamily
  astro-ph/0407196}}].

\bibitem{Das:2005yj}
S.~Das, P.~S. Corasaniti and J.~Khoury, \emph{{Super-acceleration as signature
  of dark sector interaction}},
  \href{http://dx.doi.org/10.1103/PhysRevD.73.083509}{\emph{Phys. Rev. D}
  {\bfseries 73} (2006) 083509},
  [\href{https://arxiv.org/abs/astro-ph/0510628}{{\ttfamily
  astro-ph/0510628}}].

\bibitem{Uzan:2006mf}
J.-P. Uzan, \emph{{The acceleration of the universe and the physics behind
  it}}, \href{http://dx.doi.org/10.1007/s10714-006-0385-z}{\emph{Gen. Rel.
  Grav.} {\bfseries 39} (2007) 307--342},
  [\href{https://arxiv.org/abs/astro-ph/0605313}{{\ttfamily
  astro-ph/0605313}}].

\bibitem{Uzan:2010pm}
J.-P. Uzan, \emph{{Varying Constants, Gravitation and Cosmology}},
  \href{http://dx.doi.org/10.12942/lrr-2011-2}{\emph{Living Rev. Rel.}
  {\bfseries 14} (2011) 2}, [\href{https://arxiv.org/abs/1009.5514}{{\ttfamily
  1009.5514}}].

\bibitem{Uzan:2024ded}
J.-P. Uzan, \emph{{Fundamental constants: from measurement to the universe, a
  window on gravitation and cosmology}},
  [\href{https://arxiv.org/abs/2410.07281}{{\ttfamily 2410.07281}}].

\bibitem{Ganesan:2024bsf}
V.~Ganesan, A.~Chakraborty, T.~Ray, S.~Das and A.~Banerjee, \emph{{Hint of dark
  matter-dark energy interaction in the current cosmological data?}},
  [\href{https://arxiv.org/abs/2403.14247}{{\ttfamily 2403.14247}}].

\bibitem{Khoury:2025txd}
J.~Khoury, M.-X. Lin and M.~Trodden, \emph{{Apparent $w<-1$ and a Lower $S_8$
  from Dark Axion and Dark Baryons Interactions}},
  [\href{https://arxiv.org/abs/2503.16415}{{\ttfamily 2503.16415}}].

\bibitem{Ozulker:2022slu}
E.~Ozulker, \emph{Is the dark energy equation of state parameter singular?}, \emph{Phys. Rev. D} \textbf{106} (2022) 063509, [\href{https://arxiv.org/abs/2203.04167}{{\ttfamily 2203.04167}}].

\bibitem{Chakraborty:2025syu}
A.~Chakraborty, P.~K. Chanda, S.~Das and K.~Dutta, \emph{{DESI results: Hint
  towards coupled dark matter and dark energy}},
  [\href{https://arxiv.org/abs/2503.10806}{{\ttfamily 2503.10806}}].

\bibitem{Smith:2025grk}
A.~Smith, P.~Brax, C.~van~de Bruck, C.~P. Burgess and A.-C. Davis,
  \emph{{Screened Axio-dilaton Cosmology: Novel Forms of Early Dark Energy}},
  [\href{https://arxiv.org/abs/2505.05450}{{\ttfamily 2505.05450}}].

\bibitem{Postolak:2025qmv}
M.~Postolak, \emph{{Non-minimally coupled scalar field dark sector of the
  universe: in-depth (Einstein frame) case study}},
  [\href{https://arxiv.org/abs/2505.07456}{{\ttfamily 2505.07456}}].

\bibitem{Carvalho:2004ty}
F.~C. Carvalho and A.~Saa, \emph{{Non-minimal coupling, exponential potentials
  and the w \ensuremath{<} -1 regime of dark energy}},
  \href{http://dx.doi.org/10.1103/PhysRevD.70.087302}{\emph{Phys. Rev. D}
  {\bfseries 70} (2004) 087302},
  [\href{https://arxiv.org/abs/astro-ph/0408013}{{\ttfamily
  astro-ph/0408013}}].

\bibitem{Martin:2005bp}
J.~Martin, C.~Schimd and J.-P. Uzan, \emph{{Testing for w\ensuremath{<} -1 in
  the solar system}},
  \href{http://dx.doi.org/10.1103/PhysRevLett.96.061303}{\emph{Phys. Rev.
  Lett.} {\bfseries 96} (2006) 061303},
  [\href{https://arxiv.org/abs/astro-ph/0510208}{{\ttfamily
  astro-ph/0510208}}].

\bibitem{Wolf:2025jed}
W.~J. Wolf, C.~Garc\'\i{}a-Garc\'\i{}a, T.~Anton and P.~G. Ferreira, \emph{{The
  Cosmological Evidence for Non-Minimal Coupling}},
  [\href{https://arxiv.org/abs/2504.07679}{{\ttfamily 2504.07679}}].

\bibitem{Silva:2025hxw}
E.~Silva, M.~A.~Sabogal, M.~S.~Souza, R.~C.~Nunes, E.~Di Valentino and S.~Kumar, \emph{New Constraints on Interacting Dark Energy from DESI DR2 BAO Observations}, [\href{https://arxiv.org/abs/2503.23225}{{\ttfamily 2503.23225}}].

\bibitem{Pettorino:2008ez}
V.~Pettorino and C.~Baccigalupi, \emph{{Coupled and Extended Quintessence:
  theoretical differences and structure formation}},
  \href{http://dx.doi.org/10.1103/PhysRevD.77.103003}{\emph{Phys. Rev. D}
  {\bfseries 77} (2008) 103003},
  [\href{https://arxiv.org/abs/0802.1086}{{\ttfamily 0802.1086}}].

\bibitem{Marchesano:2024gul}
F.~Marchesano, G.~Shiu and T.~Weigand, \emph{{The Standard Model from String
  Theory: What Have We Learned?}},
  \href{http://dx.doi.org/10.1146/annurev-nucl-102622-012235}{\emph{Ann. Rev.
  Nucl. Part. Sci.} {\bfseries 74} (2024) 113--140},
  [\href{https://arxiv.org/abs/2401.01939}{{\ttfamily 2401.01939}}].

\bibitem{Ooguri:2006in}
H.~Ooguri and C.~Vafa, \emph{{On the Geometry of the String Landscape and the
  Swampland}},
  \href{http://dx.doi.org/10.1016/j.nuclphysb.2006.10.033}{\emph{Nucl. Phys. B}
  {\bfseries 766} (2007) 21--33},
  [\href{https://arxiv.org/abs/hep-th/0605264}{{\ttfamily hep-th/0605264}}].

\bibitem{Grimm:2018ohb}
T.~W. Grimm, E.~Palti and I.~Valenzuela, \emph{{Infinite Distances in Field
  Space and Massless Towers of States}},
  \href{http://dx.doi.org/10.1007/JHEP08(2018)143}{\emph{JHEP} {\bfseries 08}
  (2018) 143}, [\href{https://arxiv.org/abs/1802.08264}{{\ttfamily
  1802.08264}}].

\bibitem{Andriot:2020lea}
D.~Andriot, N.~Cribiori and D.~Erkinger, \emph{{The web of swampland
  conjectures and the TCC bound}},
  \href{http://dx.doi.org/10.1007/JHEP07(2020)162}{\emph{JHEP} {\bfseries 07}
  (2020) 162}, [\href{https://arxiv.org/abs/2004.00030}{{\ttfamily
  2004.00030}}].

\bibitem{Agrawal:2019dlm}
P.~Agrawal, G.~Obied and C.~Vafa, \emph{{$H_0$ tension, swampland conjectures,
  and the epoch of fading dark matter}},
  \href{http://dx.doi.org/10.1103/PhysRevD.103.043523}{\emph{Phys. Rev. D}
  {\bfseries 103} (2021) 043523},
  [\href{https://arxiv.org/abs/1906.08261}{{\ttfamily 1906.08261}}].

\bibitem{Gonzalo:2022jac}
E.~Gonzalo, M.~Montero, G.~Obied and C.~Vafa, \emph{{Dark dimension gravitons
  as dark matter}},
  \href{http://dx.doi.org/10.1007/JHEP11(2023)109}{\emph{JHEP} {\bfseries 11}
  (2023) 109}, [\href{https://arxiv.org/abs/2209.09249}{{\ttfamily
  2209.09249}}].

\bibitem{Casas:2024oak}
G.~F. Casas and I.~Ruiz, \emph{{Cosmology of light towers and swampland
  constraints}}, \href{http://dx.doi.org/10.1007/JHEP12(2024)193}{\emph{JHEP}
  {\bfseries 12} (2024) 193},
  [\href{https://arxiv.org/abs/2409.08317}{{\ttfamily 2409.08317}}].

\bibitem{Andriot:2023wvg}
D.~Andriot, D.~Tsimpis and T.~Wrase, \emph{{Accelerated expansion of an open
  universe and string theory realizations}},
  \href{http://dx.doi.org/10.1103/PhysRevD.108.123515}{\emph{Phys. Rev. D}
  {\bfseries 108} (2023) 123515},
  [\href{https://arxiv.org/abs/2309.03938}{{\ttfamily 2309.03938}}].

\bibitem{Andriot:2025cyi}
D.~Andriot, N.~Cribiori and T.~Van~Riet, \emph{{Scale separation, rolling
  solutions and entropy bounds}},
  [\href{https://arxiv.org/abs/2504.08634}{{\ttfamily 2504.08634}}].

\bibitem{DESI:2024aqx}
{\scshape DESI} collaboration, R.~Calderon et~al., \emph{{DESI 2024:
  reconstructing dark energy using crossing statistics with DESI DR1 BAO
  data}}, \href{http://dx.doi.org/10.1088/1475-7516/2024/10/048}{\emph{JCAP}
  {\bfseries 10} (2024) 048},
  [\href{https://arxiv.org/abs/2405.04216}{{\ttfamily 2405.04216}}].

\bibitem{Kamionkowski:2022pkx}
M.~Kamionkowski and A.~G. Riess, \emph{{The Hubble Tension and Early Dark
  Energy}},
  \href{http://dx.doi.org/10.1146/annurev-nucl-111422-024107}{\emph{Ann. Rev.
  Nucl. Part. Sci.} {\bfseries 73} (2023) 153--180},
  [\href{https://arxiv.org/abs/2211.04492}{{\ttfamily 2211.04492}}].

\bibitem{Poulin:2023lkg}
V.~Poulin, T.~L. Smith and T.~Karwal, \emph{{The Ups and Downs of Early Dark
  Energy solutions to the Hubble tension: A review of models, hints and
  constraints circa 2023}},
  \href{http://dx.doi.org/10.1016/j.dark.2023.101348}{\emph{Phys. Dark Univ.}
  {\bfseries 42} (2023) 101348},
  [\href{https://arxiv.org/abs/2302.09032}{{\ttfamily 2302.09032}}].

\bibitem{Vagnozzi:2023nrq}
S.~Vagnozzi, \emph{Seven Hints That Early-Time New Physics Alone Is Not Sufficient to Solve the Hubble Tension}, \emph{Universe} \textbf{9} (2023) 393, [\href{https://arxiv.org/abs/2308.16628}{{\ttfamily 2308.16628}}].

\bibitem{Shlivko:2024llw}
D.~Shlivko and P.~J. Steinhardt, \emph{{Assessing observational constraints on
  dark energy}},
  \href{http://dx.doi.org/10.1016/j.physletb.2024.138826}{\emph{Phys. Lett. B}
  {\bfseries 855} (2024) 138826},
  [\href{https://arxiv.org/abs/2405.03933}{{\ttfamily 2405.03933}}].

\end{thebibliography}
\end{document}